\documentclass[a4paper,fleqn,numbers]{cas-sc}

\providecommand{\printemails}{}
\providecommand{\printurls}{}
\providecommand{\printorcid}{}
\providecommand{\printfacebook}{}
\providecommand{\printtwitter}{}
\providecommand{\printlinkedin}{}
\providecommand{\printgplus}{}
\usepackage{subfigure}
\usepackage{caption}
\usepackage{amsmath} 
\usepackage{natbib}
\usepackage{xcolor}
\usepackage{soul}
\usepackage{tcolorbox}

\begin{document}
\let\WriteBookmarks\relax
\def\floatpagepagefraction{1}
\def\textpagefraction{.001}

\title [mode = title]{Siamese Neural Network for Label-Efficient Critical Phenomena Prediction in 3D Percolation Models}  

\shorttitle{}    

\shortauthors{Wang S. et al.}

\author[1]{Shanshan Wang}

\affiliation[1]{organization={Key Laboratory of Quark and Lepton Physics (MOE) and Institute of Particle Physics, Central China Normal University},
            addressline={},
            city={Wuhan},
            postcode={430079},
            state={Hubei},
            country={China}}

\author[1,2]{Dian Xu}

\affiliation[2]{organization={School of Engineering and Technology, Baoshan University},
            addressline={},
            city={Baoshan},
            postcode={678000},
            state={Yunnan},
            country={China}}

\author[1,2]{Jianmin Shen}

\author[1]{Feng Gao}

\author[1,3]{Wei Li}

\author[1]{Weibing Deng}
\cormark[1]
\ead{wdeng@mail.ccnu.edu.cn}

\affiliation[3]{organization={Ecole Superieure d'Informatique Electronique Automatique},
            addressline={},
            city={Ivry-sur-Seine},
            postcode={94200},
            state={},
            country={France}}

\cortext[1]{Corresponding author}

\begin{abstract}
Predicting critical phenomena from limited labeled data remains a challenging task in statistical physics. As percolation theory provides a canonical model for phase transitions with well-established critical exponents, it serves as an ideal benchmark for validating new machine learning frameworks.
Here, we introduce a label-efficient learning framework based on a Siamese Neural Network (SNN) to identify phase transitions in three-dimensional site and bond percolation models. Using only 22 labeled probability points drawn entirely from non-critical regions, the method locates percolation thresholds with percent-level accuracy and yields estimates of the critical exponent $\nu$ consistent with literature values within statistical uncertainty. Analysis of the learned representations clarifies what the network learns: although trained solely on binary similarity labels, the network autonomously converges to a statistic that coincides quantitatively with the normalized largest-cluster size $S_{max}/L^3$ ($r > 0.99$), the finite-size order parameter of percolation. This underlies the framework's most distinctive capability---a model trained solely on simple cubic lattices identifies the phase transition in face-centered cubic lattices without retraining. The framework thus offers a complementary route to criticality detection in settings where no quantitative order parameter is explicitly defined or labeled data is scarce.
\end{abstract}


\maketitle

\section{Introduction}
The advent of machine learning (ML) has profoundly influenced scientific inquiry by enabling powerful modeling of complex systems across scales, from quantum materials to astrophysical networks\cite{alzubi2018machine, thiyagalingam2022scientific}. Beyond its transformative impact on industrial paradigms—evident in autonomous systems\cite{shalev2016safe}, personalized medicine\cite{Char2018ImplementingML}, and algorithmic finance\cite{Lin2012MachineLI}—ML is reshaping fundamental physics research by addressing long-standing challenges in statistical mechanics and phase transition theory\cite{Jordan2015MachineLT, mohri2018foundations}. Specifically, ML methodologies offer a paradigm shift by transcending the limitations of traditional analytical frameworks, enabling novel insights into emergent phenomena, and simultaneously complementing conventional computational and experimental methods\cite{xue2023machine, hu2023universality}. This synergy between data-centric algorithms and physical principles not only enhances predictive accuracy but also unveils hidden patterns in complex systems, highlighting the potential of ML as a powerful tool for modern theoretical physics.

Phase transitions, as cornerstones of statistical and condensed matter physics, have traditionally been analyzed through theoretical frameworks and numerical simulations~\cite{yangyuxiang, lifshitz2013statistical}. Monte Carlo (MC) methods have achieved high precision in studying critical phenomena across dimensions. From an algorithmic perspective, the study of percolation has historically been driven by major computational breakthroughs. Early approaches relied on graph traversal techniques, such as depth-first search (DFS)~\cite{tarjan1972depth}, to systematically identify connected components. A significant leap occurred with the Hoshen-Kopelman algorithm~\cite{hoshen1976percolation}, which utilized union-find data structures to enable efficient single-pass cluster labeling. Subsequently, the Newman-Ziff algorithm~\cite{newman2000efficient, newman2001fast} revolutionized the field by employing microcanonical sampling, allowing for the analysis of the entire probability range in a single simulation run.

Driven by these algorithmic innovations, the precision of percolation thresholds has steadily improved over the decades. Modern high-precision simulations have established the thresholds for three-dimensional site and bond percolation as $p_c = 0.311 607 68(15)$ and $p_c = 0.248 811 85(10)$, respectively~\cite{xu2014simultaneous}. Critical exponents have also been determined to high accuracy, such as $\nu \approx 0.8762(7)$~\cite{brzeski2022percolation}. It is worth noting that the prediction accuracy achieved by our SNN framework is comparable to early representative studies, such as those by Sur \textit{et al.} ($p_c = 0.3115(5)$)~\cite{sur1976monte} and Sykes and Essam ($p_c = 0.247(5)$)~\cite{sykes1964critical}. This positions our current work not as a competitor to modern high-precision algorithms, but as a methodological proof-of-concept, demonstrating that neural networks can autonomously learn topological phase transitions with accuracy levels consistent with early numerical experiments.

Complementing these traditional approaches, recent research has increasingly integrated machine learning, particularly deep learning architectures, into phase transition studies. Contemporary research demonstrates ML's transformative potential in predicting critical phenomena and elucidating nonequilibrium dynamics~\cite{hu2017discovering, van2017learning}. A landmark study showed that a neural network can learn to distinguish different phases in the 2D Ising gauge model directly from raw spin configurations, demonstrating that ML methods can automatically extract physically meaningful features without requiring prior knowledge of order parameters~\cite{Carrasquilla2016MachineLP}. While percolation theory finds broad applications in critical phenomena and material science~\cite{essam1980percolation, pike1974percolation}, the application of ML to these well-understood systems offers a unique opportunity: it serves as an ideal benchmark for validating the ability of neural networks to extract topological features from complex configurations. By learning low-dimensional representations, ML offers a scalable approach to criticality prediction, particularly for systems where the order parameter is unknown or experimental data is scarce.

Recent advancements in ML have catalyzed a paradigm shift in statistical physics research, particularly in the analysis of phase transitions across supervised, unsupervised, and semi-supervised learning frameworks:

1) Supervised learning for phase discrimination. The synergy between deep learning and RG theory has enabled breakthroughs in identifying topological phase transitions. Convolutional Neural Networks (CNNs) have been explored in the 2D Ising model using RG-inspired architectures, illustrating that ML can capture essential features of phase boundaries from raw configurations\cite{Mehta2018AHL}. This approach was extended to the XY model, where supervised classification successfully captured both the Berezinskii-Kosterlitz-Thouless (BKT) transition and percolation threshold behaviors by encoding spin configurations as input features\cite{PhysRevE.99.032142}. The methodology was further extended to 3D Ising systems, demonstrating CNNs’ capacity to generalize across dimensionalities while producing results consistent with state-of-the-art Monte Carlo studies\cite{Li2021MachineLP}.

2) Unsupervised learning for label-free phase detection. The difficulty of defining or obtaining labeled data in certain many-body systems has motivated the use of unsupervised learning techniques, which aim to uncover phase structure and critical behavior without explicit supervision. Siamese Neural Networks have enabled unsupervised identification of phase boundaries in both Monte Carlo simulations and Rydberg atom arrays, revealing multiple phases without prior labeling or knowledge of their existence\cite{patel2022unsupervised}. Principal component analysis (PCA) has enabled dimensionality reduction for Ising model phase classification, revealing hidden structures in magnetization landscapes\cite{Wang2016DiscoveringPT}. A notable advancement is the use of unsupervised learning methods—such as principal component analysis and variational autoencoders—for identifying phase transitions in Ising and XY models without prior knowledge of the Hamiltonian or order parameter. These approaches reveal clustered latent representations that correspond to distinct phases, enabling label-free detection of critical behavior\cite{Wetzel2017UnsupervisedLO}.

3) Semi-supervised learning for phase transitions.
Semi-supervised approaches, particularly Domain Adversarial Neural Networks (DANNs), have emerged as pivotal tools for transfer learning in phase transition studies. Notably, DANNs have been successfully applied to both equilibrium and nonequilibrium models—specifically, 2D percolation and directed percolation—demonstrating that reliable critical point prediction can be achieved with only a small, automatically selected set of labeled configurations\cite{shenTransferLearningPhase2022}. This framework’s robustness was validated in 3D Potts models with varying q-state symmetries, where data collapse techniques confirmed that DANN accurately captured critical behavior and generalized across universality classes\cite{Chen2022StudyOP, chenApplicationsDomainAdversarial2024}.

Despite these advancements, traditional supervised classifiers often require manually engineered order parameters and large labeled datasets, which limits their practicality when prior physical knowledge is unavailable or data annotation is costly. To address these limitations, we introduce a SNN-based approach for phase transition prediction in 3D percolation models. Unlike CNNs, which rely on extensive training sets, SNNs apply metric learning to map percolation configurations into a latent representation space, where the learned distances reflect phase-related similarity as defined during training. In the learned representation space, configurations from the same phase tend to cluster together, while near-critical configurations exhibit intermediate features and often reside near decision boundaries, enabling precise localization of the phase transition.

The key contributions of this work are:

(1) Label-Efficient Learning from Non-Critical Regions. By training on small labeled subsets sampled exclusively from non-critical regions—parameter regimes far from the percolation threshold (e.g., 1000 configurations at $p \in [0, 0.1] \cup [0.9, 1]$) across multiple system sizes, SNNs can infer critical behavior near the transition point. Our systematic experiments on labeling intervals confirm that extending the labeled range toward criticality yields no significant improvement, demonstrating that reliable threshold estimation is achievable with only 22 labeled probability points and without any supervision near the critical region.

(2) Structure-Aware Learning with Autonomous Order-Parameter Discovery. We treat the DFS-extracted largest cluster—a generic, parameter-free structural object—as input to the SNN. The division of labor is deliberate: the preprocessing specifies only which structure to attend to (connectivity), while the network autonomously discovers which function of that structure discriminates the phases. Our analysis of the learned representations (Section~\ref{sec46}) shows that the dominant direction of the embedding space coincides quantitatively with the order parameter $S_{max}/L^3$ ($r > 0.99$), although no quantitative observable was ever supplied during training. A raw-input ablation delineates the boundary of this capability: without the structural prior, the network instead learns the trivial occupation density and fails to locate the transition.

(3) Cross-Geometry Transferability. A model trained solely on simple cubic (SC) lattices ($z=6$) identifies the phase transition in face-centered cubic (FCC) lattices ($z=12$) without any retraining—a form of transferability not available within the conventional Monte Carlo paradigm, where each new geometry requires a separate simulation campaign. This zero-shot transfer is mechanistically grounded in contribution (2): the learned representation encodes an intrinsic property of the cluster—its mass—rather than lattice-specific geometry. Additionally, successful retraining at larger system sizes ($L=64$) confirms the scalability of the training protocol.

It should be emphasized that the proposed SNN-based framework is not intended to replace conventional Monte Carlo simulations or finite-size scaling analyses—which remain the standard for high-precision threshold determination on regular lattices—but to complement them by providing a data and label-efficient route for identifying criticality directly from configuration data, particularly in settings where no quantitative order parameter is explicitly defined or where labeled data is scarce.

The remainder of this paper is organized as follows: Section 2 formalizes the 3D percolation model and finite-size scaling (FSS) relations. Section 3 details the SNN architecture, contrastive learning formulation, and DFS-based feature extraction.
Section 4 presents experimental results, including ablation studies on labeled data size, labeling interval extensions, and evaluation across site and bond percolation models.
Section 5 concludes with implications for high-throughput materials discovery and open challenges in few-shot phase transition learning.

\section{Percolation Models}

The percolation model, originally developed to investigate fluid flow through porous media such as coal seams, has since become a foundational framework in modern network theory, particularly in the study of resilience, epidemic spreading, and connectivity in complex networks\cite{stauffer2018introduction,shante1971introduction}. In this context, it describes the stochastic addition or removal of nodes and edges, thereby capturing topological transitions in complex systems\cite{saberi2015recent}. Building on this versatility, the theoretical framework of percolation has been progressively extended to more generalized topologies. For instance, Shang analyzed inhomogeneous long-range percolation on hierarchical lattices\cite{shang2015inhomogeneous}, explored $k$-core percolation resilience on multiplex networks\cite{shang2020generalized}, and investigated the impact of anchor nodes on phase transitions in random hypergraphs\cite{shang2025percolation}. These works highlight the diverse critical behaviors emerging in complex structural settings. Within statistical physics, percolation serves as a paradigmatic model for analyzing phase transitions and critical phenomena, with applications spanning condensed matter physics, materials science, and the dynamics of complex networks\cite{isichenko1992percolation,araujo2014recent}.

In a lattice system, each site (or bond) exists in one of two states: occupied or vacant. Such configurations constitute random disordered systems. When adjacent occupied sites form a connected component via nearest-neighbor adjacency, they constitute a cluster. As the occupation probability \( p \) increases, the system transitions from a fragmented state of isolated clusters to a connected state where at least one macroscopic cluster spans the entire lattice. This transition occurs at the critical occupation probability \( p_c \), known as the percolation threshold.

The critical behavior of the system is characterized by the order parameter \( P_\infty(p) \), defined as the probability that a randomly selected site (or bond) belongs to the infinite percolation cluster at occupation probability \( p \). Near \( p_c \), \( P_\infty(p) \) exhibits a power-law singularity:
\begin{equation}
P_\infty(p) \propto (p - p_c)^\beta, \quad \text{for} \quad p \to p_c^+,
\label{Pp}
\end{equation}
\noindent where $\beta$ is the critical exponent that governs the scaling behavior of the order parameter near the percolation threshold. This geometrically driven transition highlights the largest cluster as a key structural feature for detecting criticality.
In finite-size simulations, the percolation (spanning) probability $P_L(p)$ is defined as the fraction of Monte Carlo realizations that contain at least one spanning cluster connecting opposite boundaries of the system.
As $L \to \infty$, $P_L(p)\!\to\!0$ for $p<p_c$ and $P_L(p)\!\to\!1$ for $p>p_c$; crossings of $P_L(p)$ for different $L$ provide a standard route to estimate $p_c$.
This quantity is distinct from the order parameter $P_\infty(p)$, which denotes the fraction of sites in the infinite cluster for $p>p_c$.

In finite-size systems, the size of the largest cluster, $S_{\text{max}}$, is a key observable for understanding critical behavior\cite{christensen2005complexity}. As the occupation probability \( p \) approaches the critical threshold \( p_c \), the size of the largest cluster \( S_{\text{max}} \) increases sharply due to the emergence of long-range correlations, but remains finite owing to the system's limited size. For \( p > p_c \), the largest cluster approximates the infinite percolation cluster in the thermodynamic limit. Its size in finite systems follows a FSS relation governed by the critical exponents \( \beta \) and \( \nu \):

\begin{equation}
S_{\text{max}} \sim L^{d - \frac{\beta}{\nu}} \cdot f\left( (p - p_c) L^{1/\nu} \right),
\end{equation}

\noindent where $L$ is the system size, $d$ is the spatial dimension, and $f(x)$ is a universal scaling function. Here, $\nu$ characterizes the divergence of the correlation length near \( p_c \), while \( \beta \) governs the scaling of the order parameter \( P_\infty(p) \). In this work, we extract the largest cluster at each configuration as a structured representation, which serves as the input to our similarity-based learning model for capturing critical trends.
 
\begin{figure}[H]
	\centering
	\subfigure{
		\begin{minipage}[b]{0.27\textwidth}
			\centering
			\includegraphics[width=0.9\textwidth]{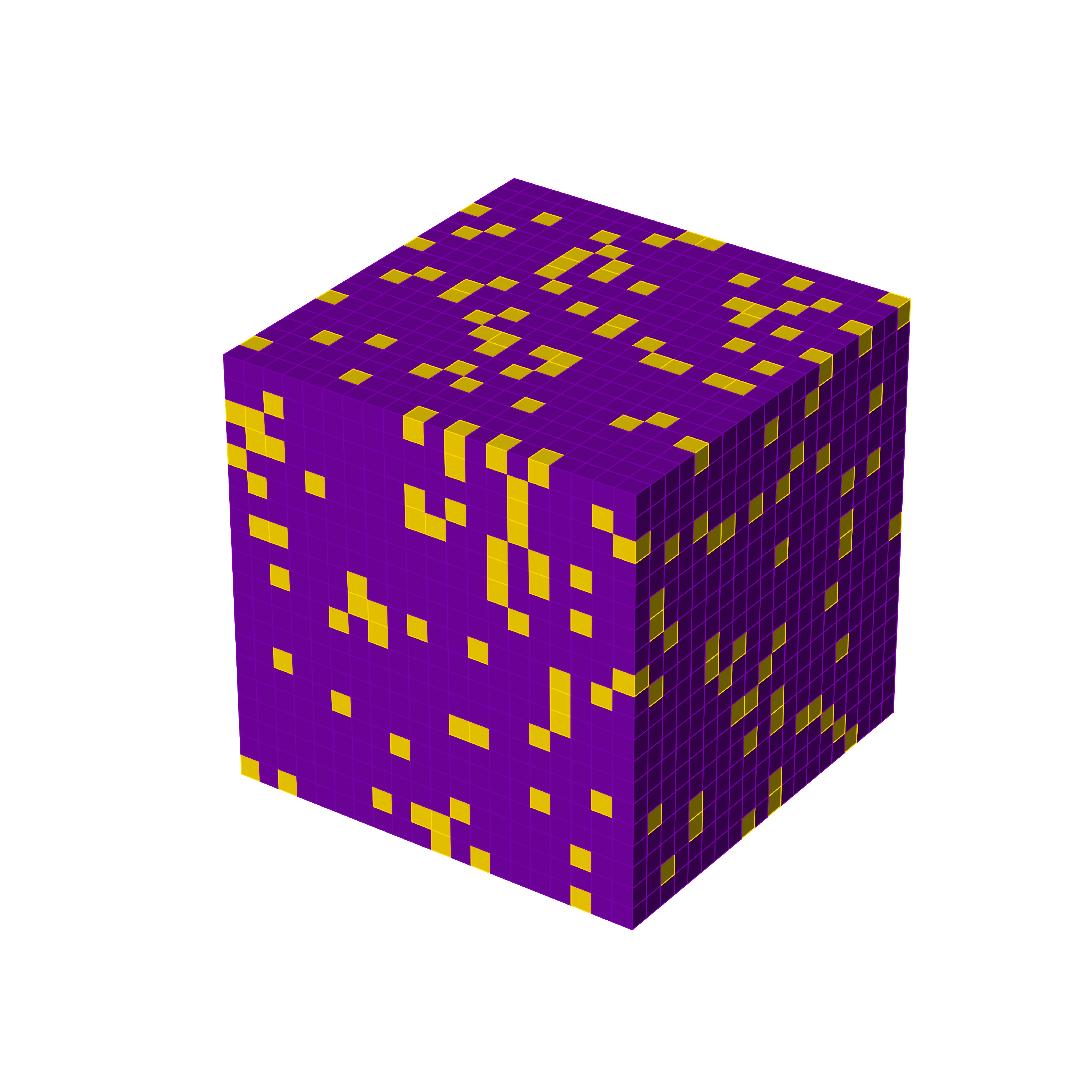} 
		\end{minipage}
	}
	\subfigure{
		\begin{minipage}[b]{0.27\textwidth}
			\centering
			\includegraphics[width=0.9\textwidth]{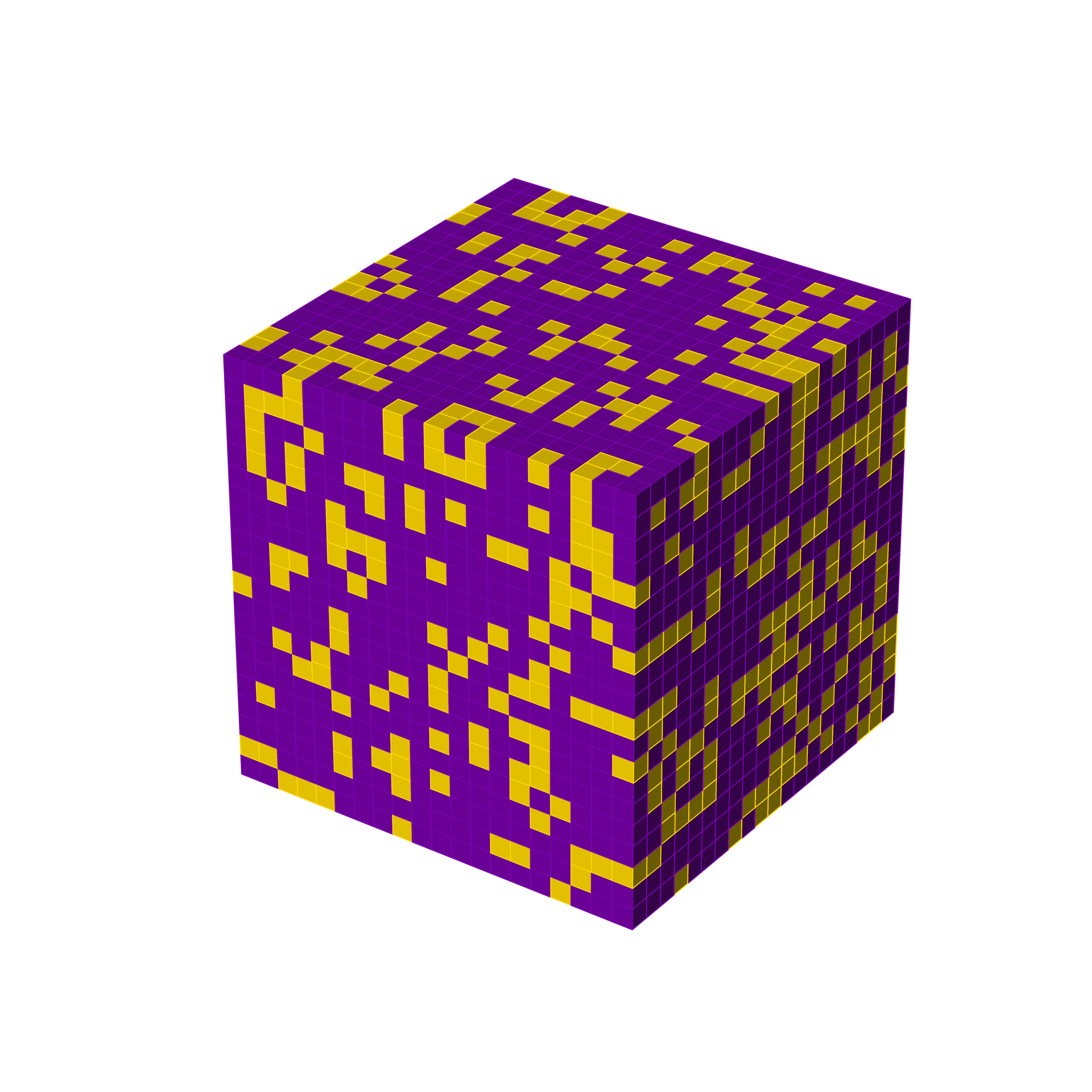}
		\end{minipage}
	}
	\subfigure{
		\begin{minipage}[b]{0.27\textwidth}
			\centering
			\includegraphics[width=0.9\textwidth]{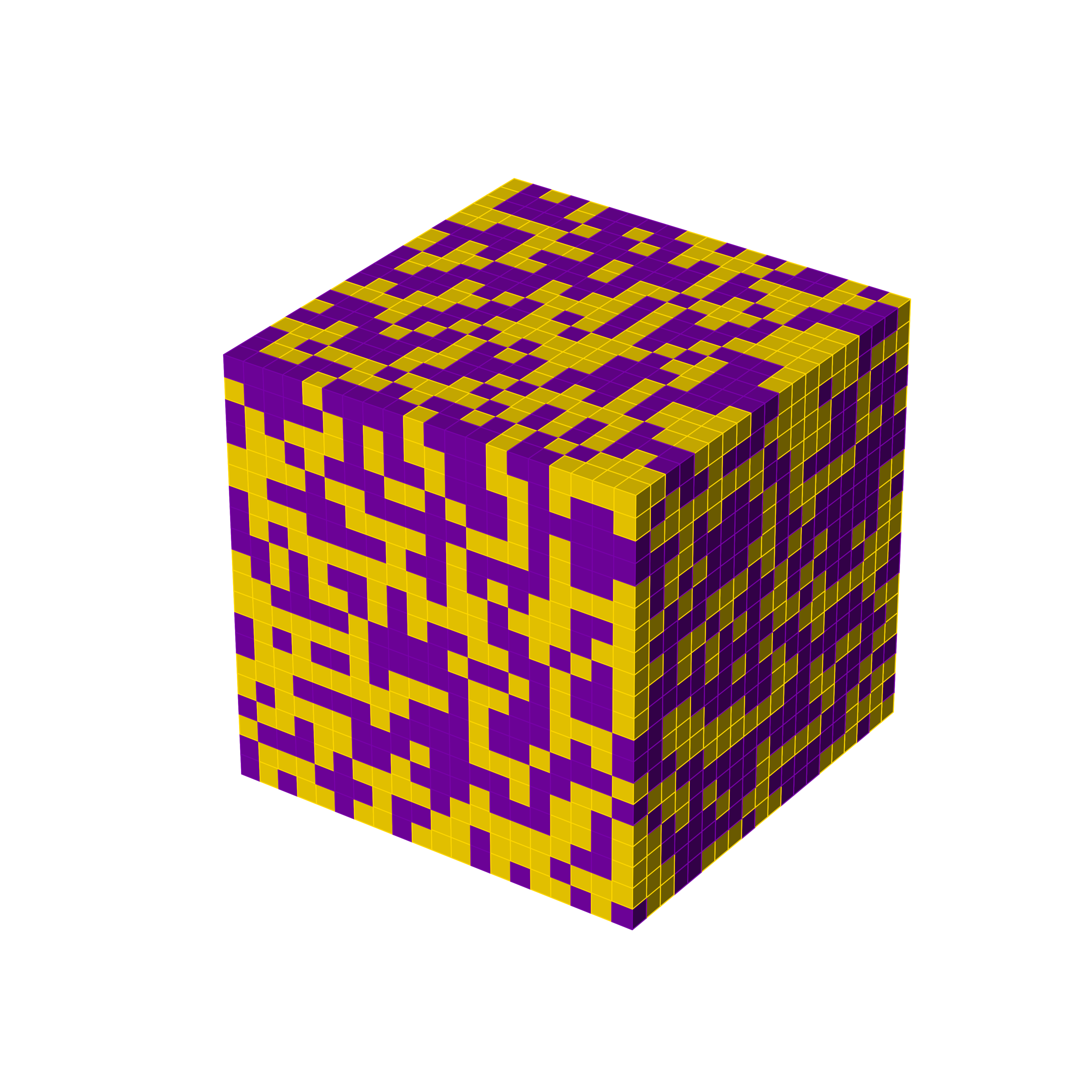}
		\end{minipage}
	}
	\\ 
	\vspace{-0.5cm}
	\subfigure{
		\begin{minipage}[b]{0.27\textwidth}
			\centering
			\includegraphics[width=0.9\textwidth]{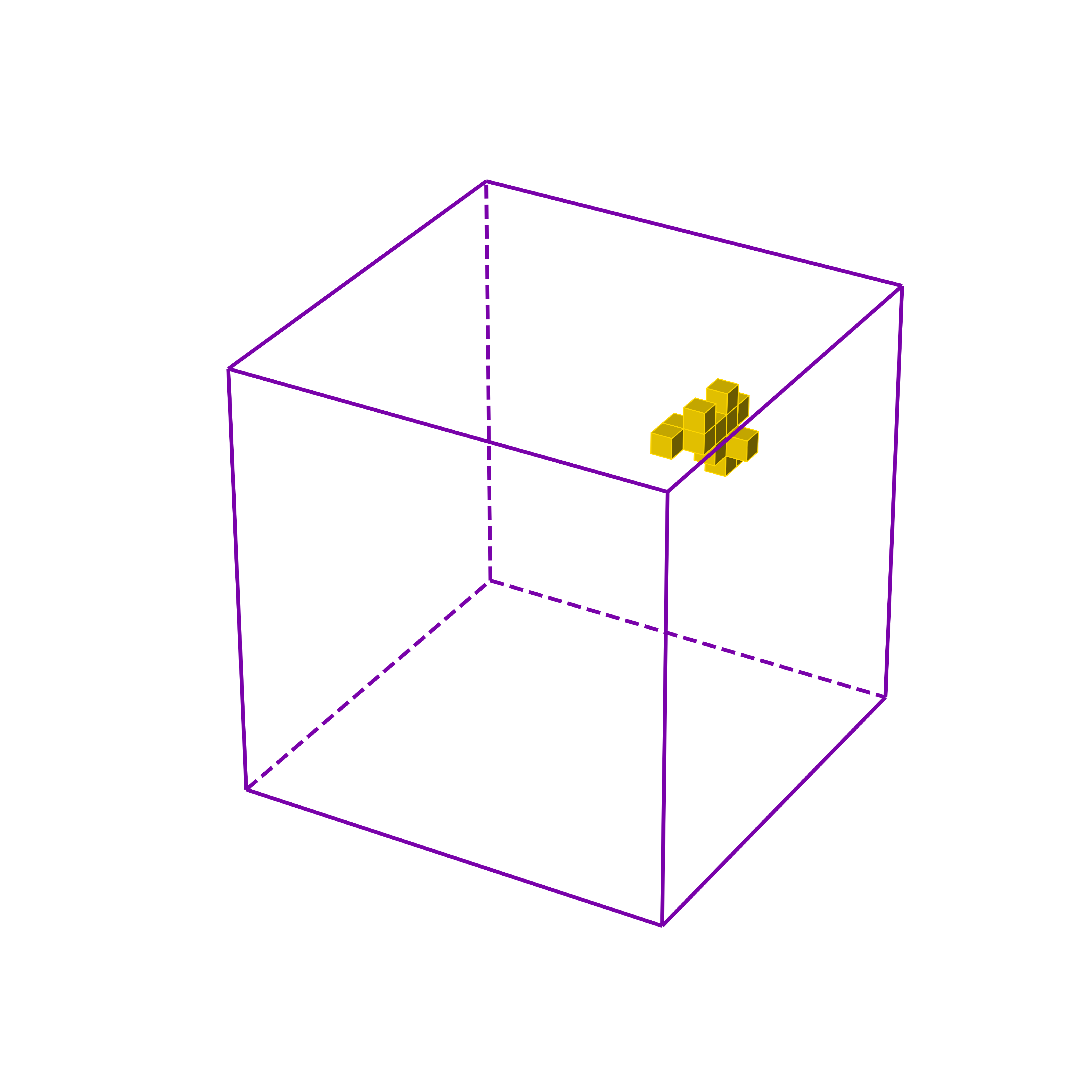} 
		\end{minipage}
	}
	\subfigure{
		\begin{minipage}[b]{0.27\textwidth}
			\centering
			\includegraphics[width=0.9\textwidth]{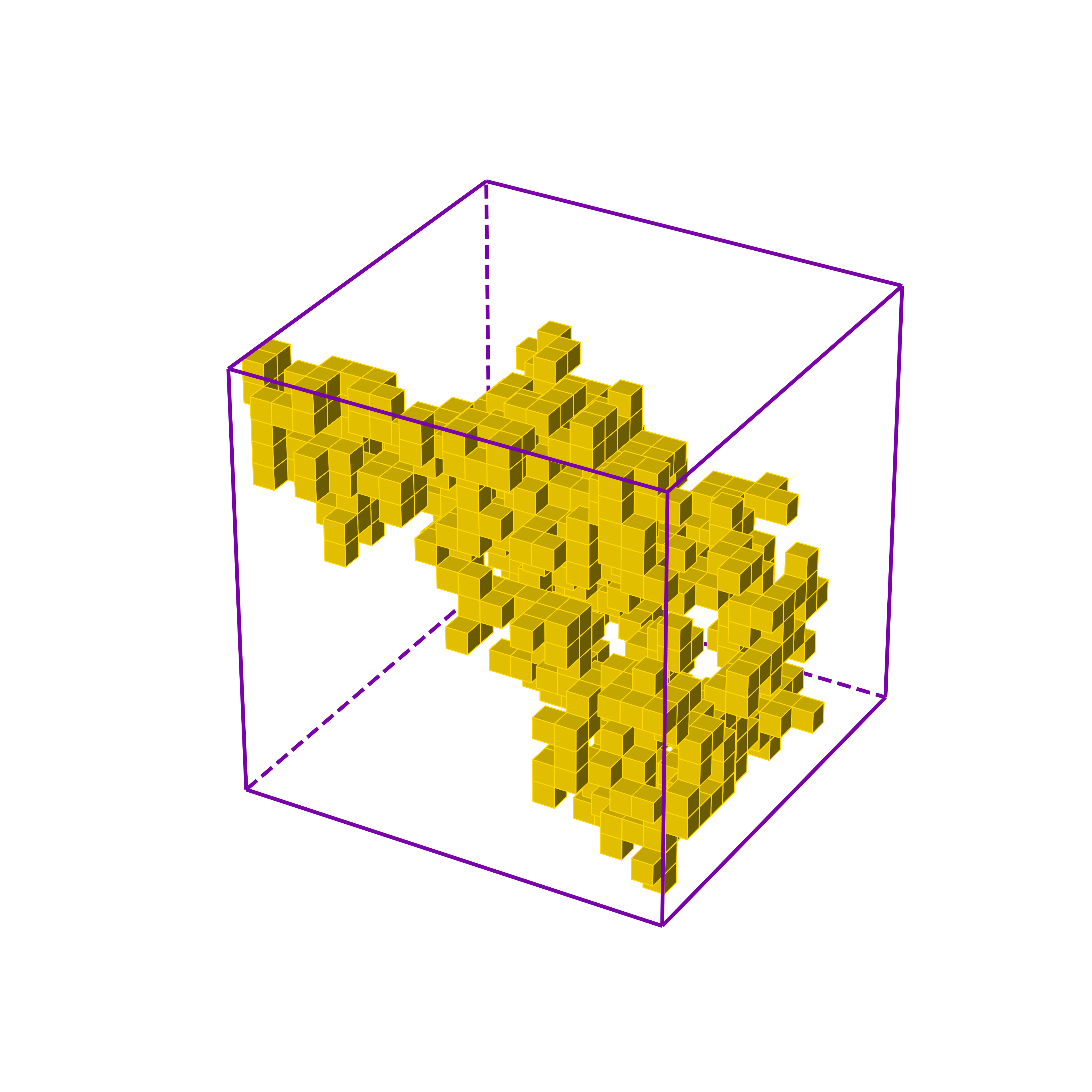}
		\end{minipage}
	}
	\subfigure{
		\begin{minipage}[b]{0.27\textwidth}
			\centering
			\includegraphics[width=0.9\textwidth]{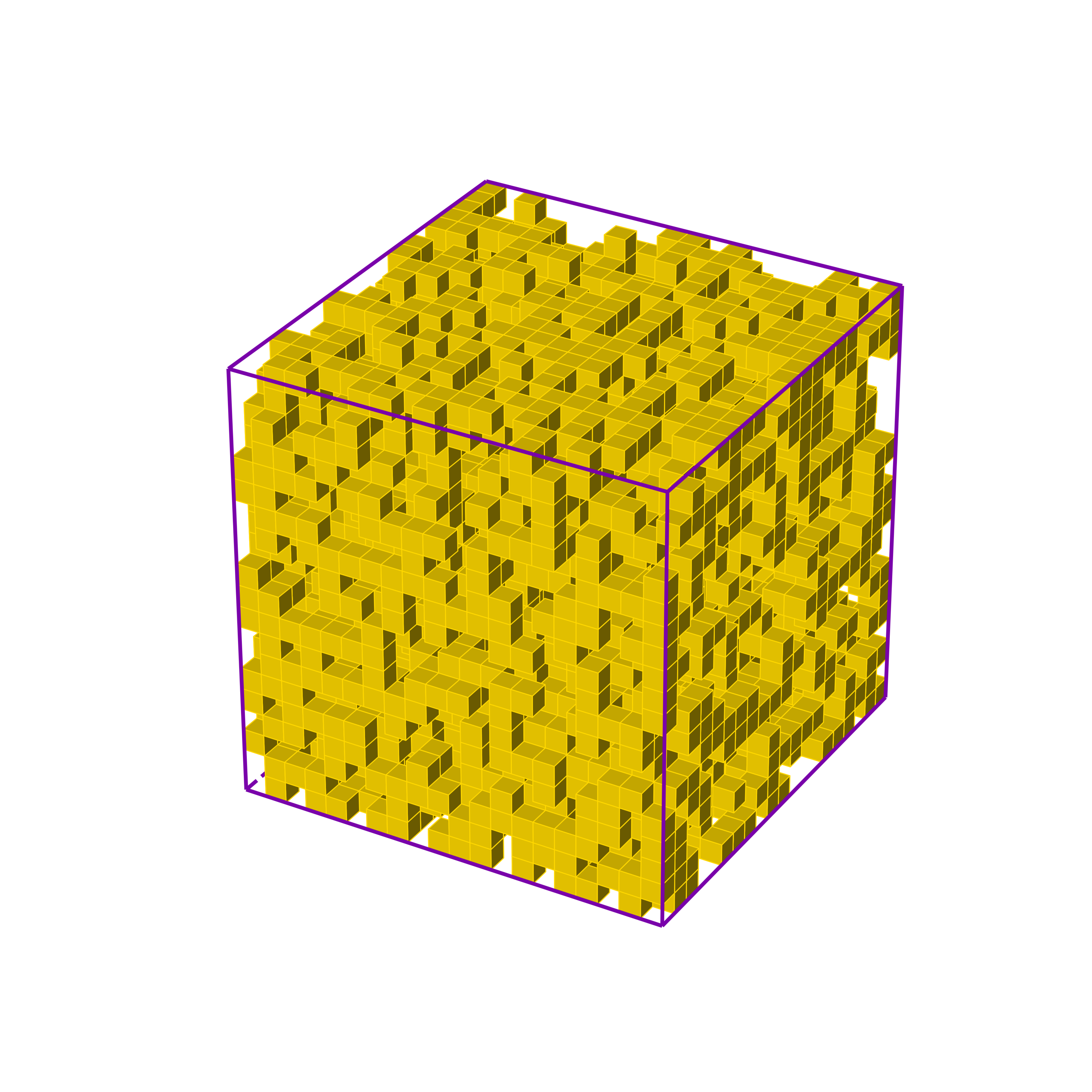}
		\end{minipage}
	}
    \caption{
    Site percolation on a three-dimensional cubic lattice. Top: Raw configurations at occupation probabilities \( p = 0.15 \), \( p = 0.312 \) (near the critical threshold \( p_c \approx 0.3116 \) for this system size), and \( p = 0.47 \). Bottom: Corresponding largest connected clusters, illustrating the transition from isolated clusters to a macroscopic spanning structure as \( p \) increases. The system size is \( L = 20 \). Occupied sites are shown; vacant sites are omitted for visual clarity.
    }
	\label{fig:3d_site_configurations}
\end{figure}
\begin{figure}[H]
	\centering
	\begin{minipage}{0.27\linewidth}
		\centering
		\includegraphics[width=0.9\linewidth]{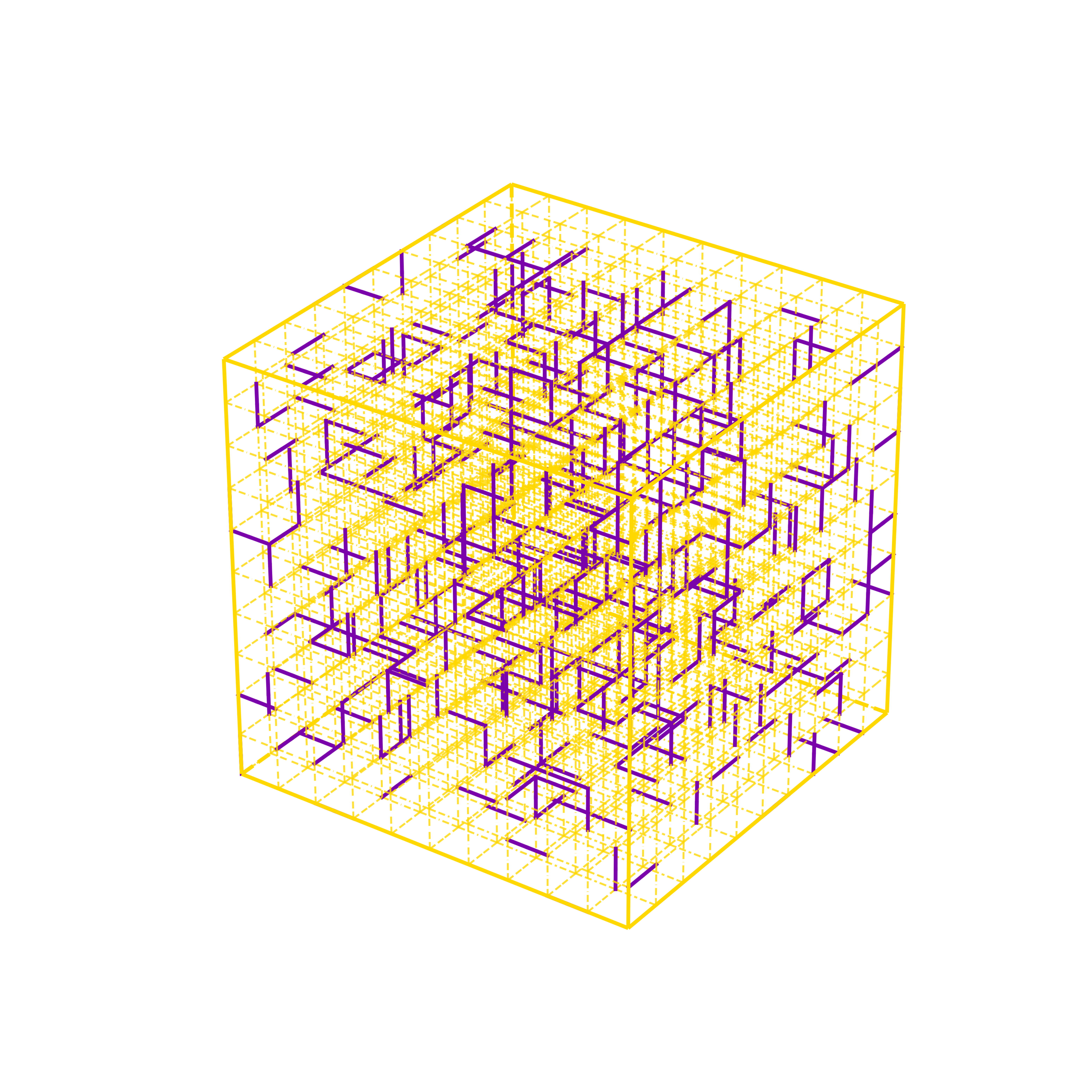}
	\end{minipage}
	\begin{minipage}{0.27\linewidth}
		\centering
		\includegraphics[width=0.9\linewidth]{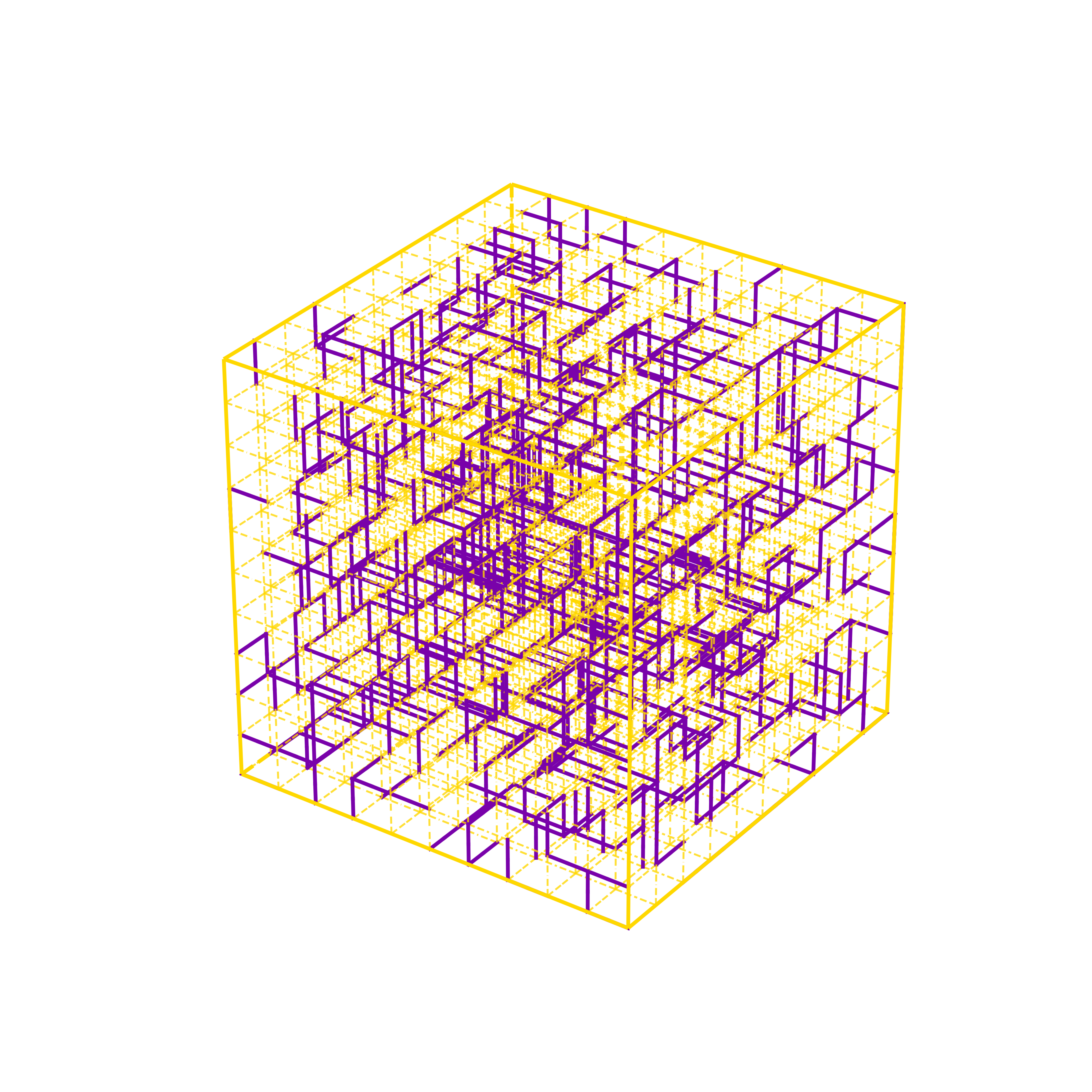}
	\end{minipage}
	\begin{minipage}{0.27\linewidth}
		\centering
		\includegraphics[width=0.9\linewidth]{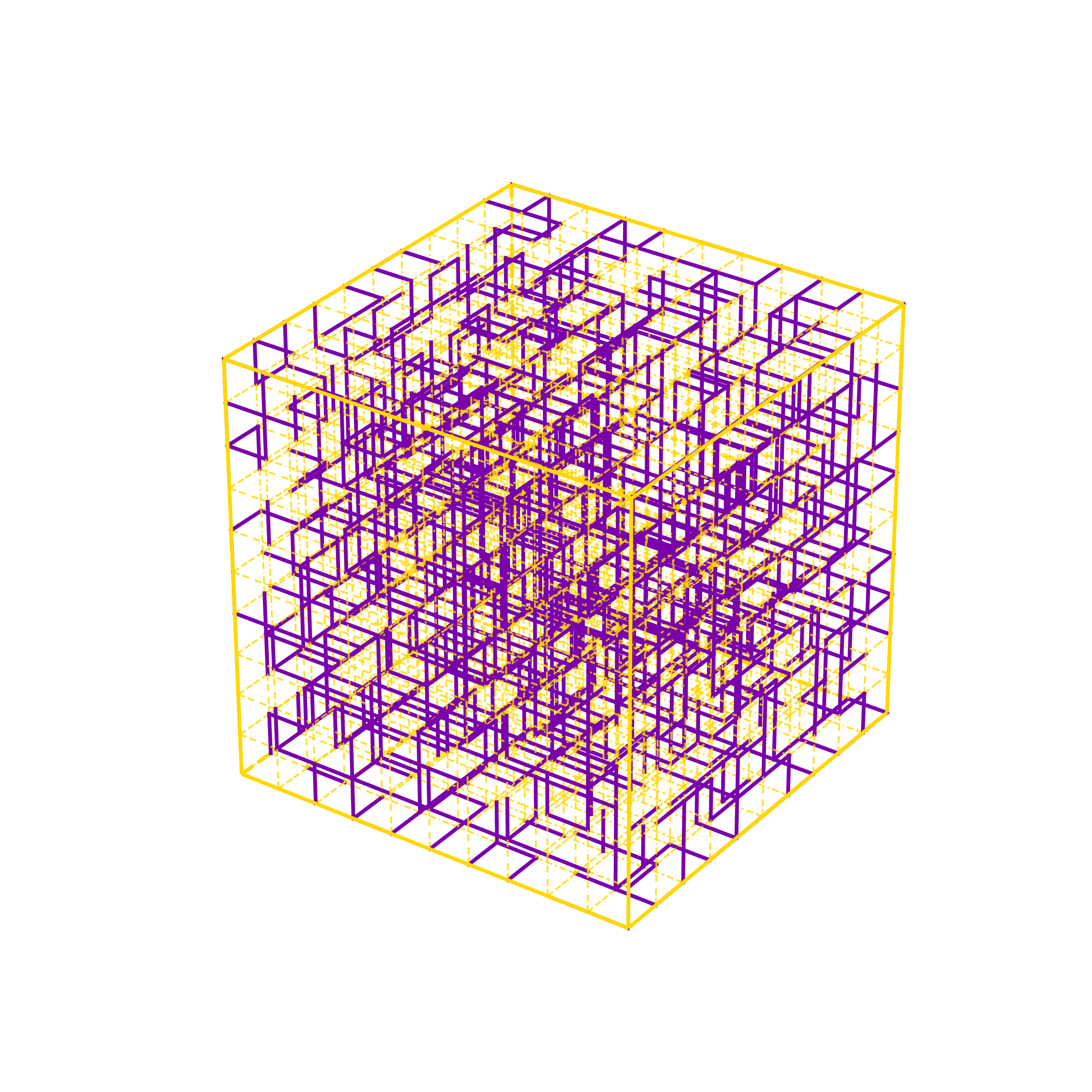}
	\end{minipage}\\
	\begin{minipage}{0.27\linewidth}
		\centering
		\includegraphics[width=0.9\linewidth]{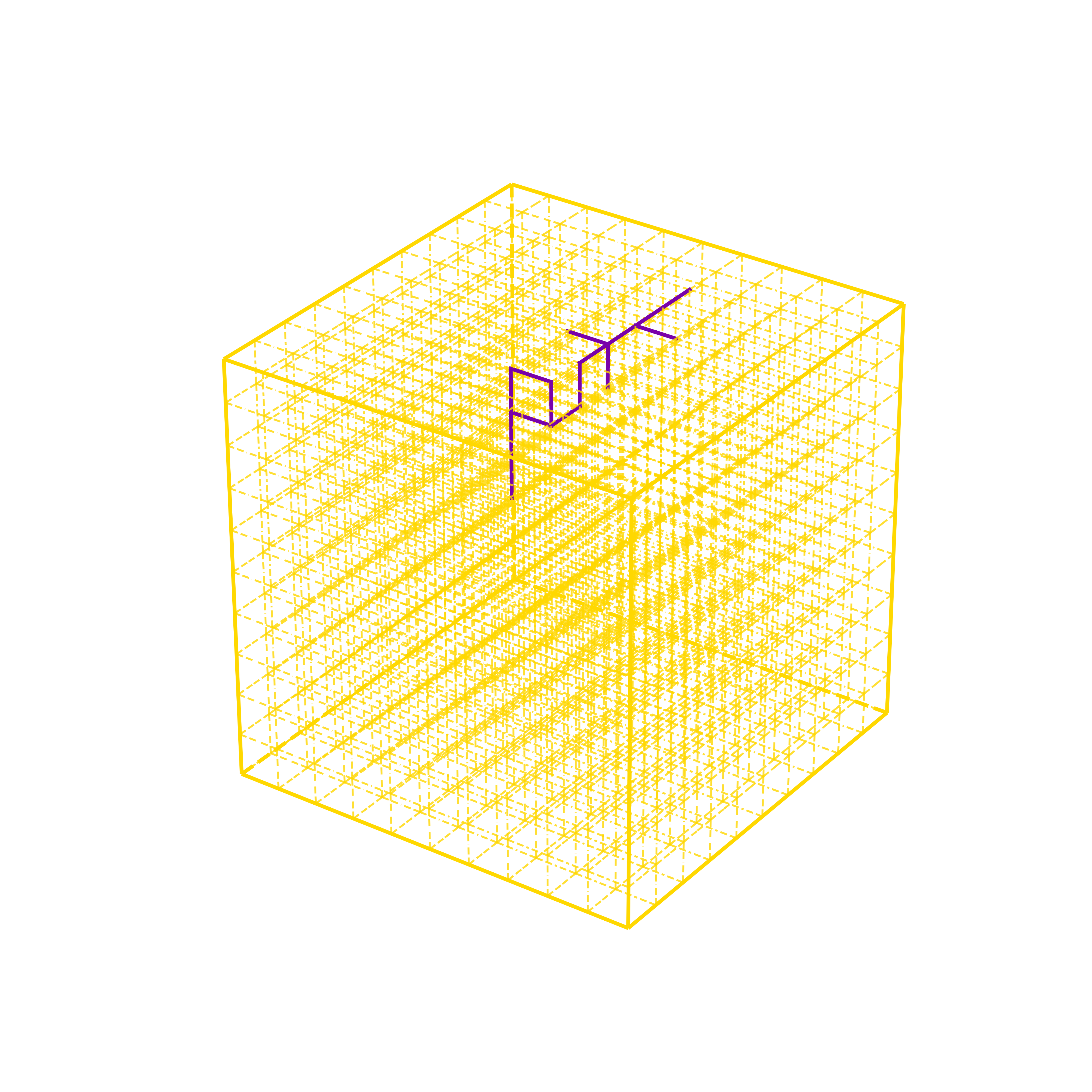}
	\end{minipage}
	\begin{minipage}{0.27\linewidth}
		\centering
		\includegraphics[width=0.9\linewidth]{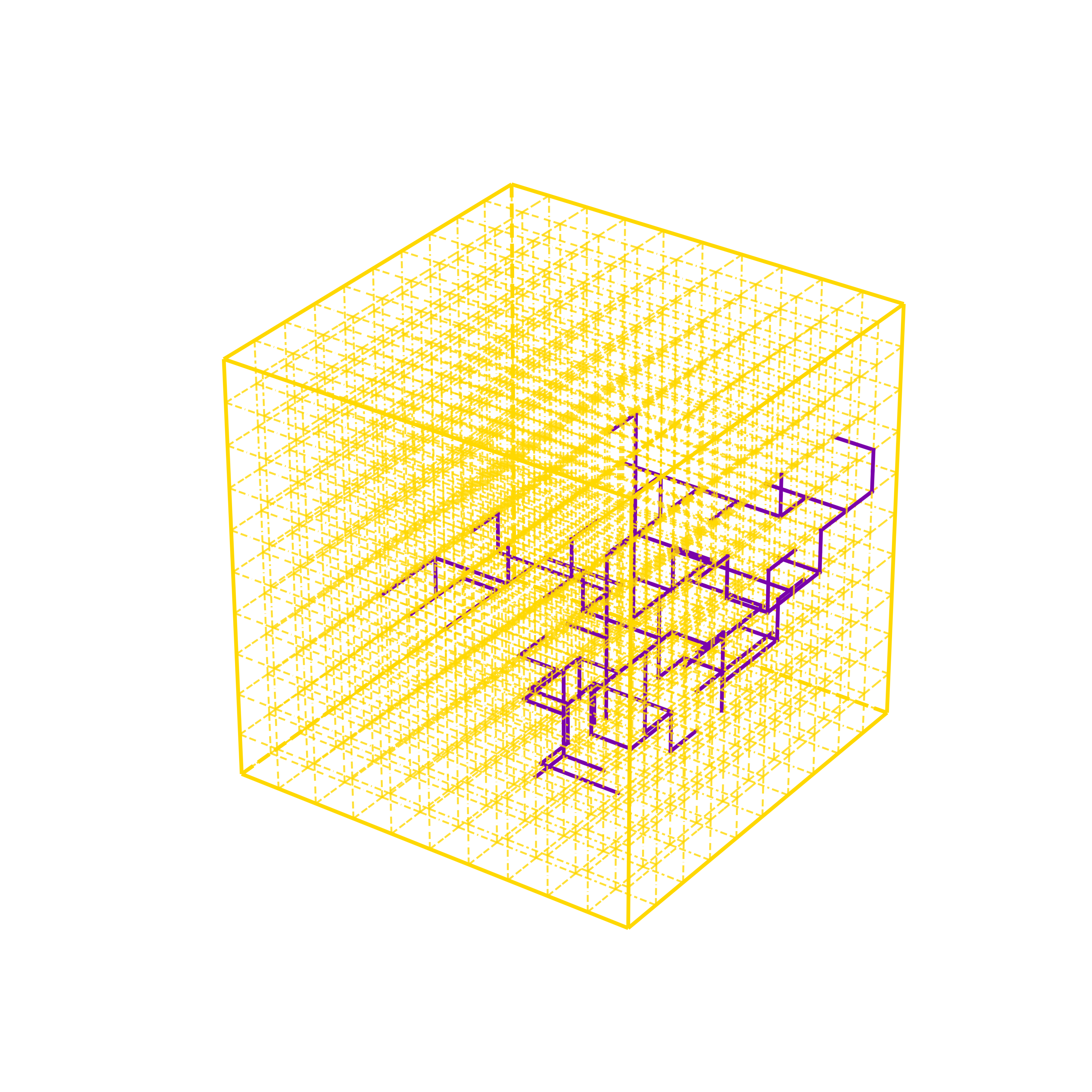}
	\end{minipage}
	\begin{minipage}{0.27\linewidth}
		\centering
		\includegraphics[width=0.9\linewidth]{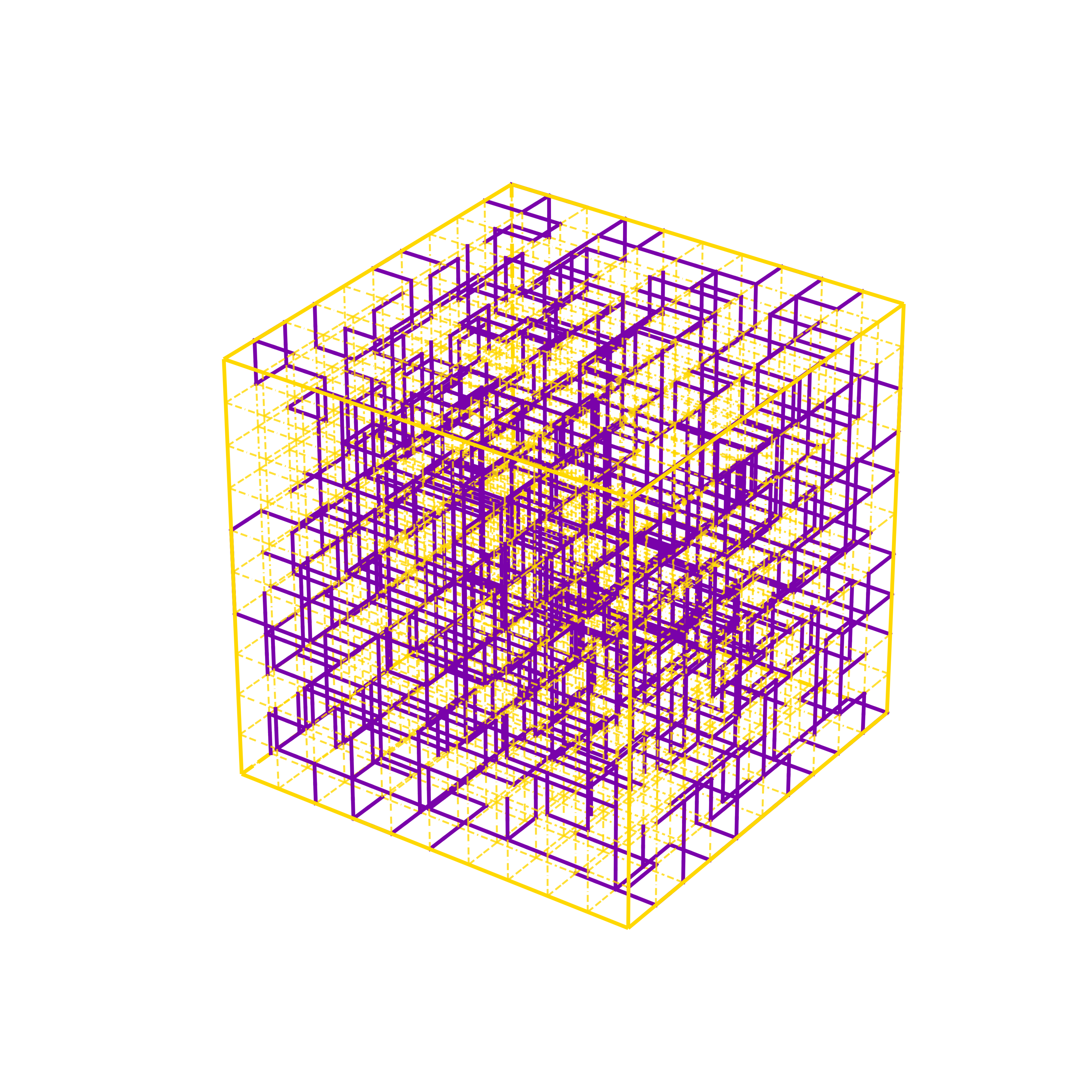}
	\end{minipage}
  \caption{Bond percolation on a three-dimensional cubic lattice. Top: Raw configurations at occupation probabilities \( p = 0.12 \), \( p = 0.249 \) (near the critical threshold \( p_c \approx 0.2488 \) for this system size), and \( p = 0.38 \). Bottom: Corresponding largest connected clusters, illustrating the emergence of a macroscopic spanning structure as \( p \) increases. The system size is \( L = 10 \). Occupied (“open”) bonds are shown in red; vacant (“closed”) bonds are omitted for visual clarity.}
	\label{fig:3d_bond_configurations}
\end{figure}
The site and bond percolation models studied in this work are defined on a regular three-dimensional cubic lattice\cite{christensen2005complexity, Fessler2007PercolationAP}. 
We choose the three-dimensional percolation model as the subject of this study not because it requires new numerical algorithms—indeed, established methods such as the Hoshen-Kopelman and Newman-Ziff algorithms serve as the gold standard for efficiency and precision in this domain—but because it provides an ideal benchmark system with known critical exponents. By testing the SNN framework on this well-understood physical model, we aim to validate its capability to learn topological features and identify phase transitions via a data-driven, metric-learning approach, thereby establishing a foundation for its future application to complex systems where standard lattice definitions may not apply.
Figures~\ref{fig:3d_site_configurations} and~\ref{fig:3d_bond_configurations} illustrate typical configurations of the site and bond percolation models on a three-dimensional cubic lattice at various occupation probabilities \( p \). These figures demonstrate that as \( p \) increases, the largest cluster grows rapidly and eventually spans the entire lattice when \( p > p_c \), signaling the emergence of a spanning cluster.

\section{Siamese Neural Network}
SNNs\cite{bromley1993signature, ranasinghe2019semantic, chicco2021siamese, zhang2016siamese} were originally introduced by Bromley in 1993 for signature verification tasks, with the primary goal of learning a similarity metric between input pairs. An SNN typically consists of two identical feedforward networks that share weights and process paired inputs in parallel. The similarity between inputs is evaluated by computing the distance between their output embeddings, using metrics such as Euclidean or cosine distance. During training, the network is optimized using a contrastive or triplet loss, which encourages similar pairs to produce nearby embeddings and dissimilar pairs to remain distant. SNNs have been widely applied in face recognition\cite{taigman2014deepface}, image retrieval\cite{koch2015siamese}, and audio representation\cite{manocha2018content}, particularly in scenarios involving limited labeled data or large category spaces. This makes SNNs particularly suitable for physical systems where configuration labeling is costly or limited. In this work, we employ a dual-input SNN to measure the similarity between configuration pairs in three-dimensional percolation models.

\vspace{0.3em}
{\bf 1) DFS-Based Largest Cluster Extraction for Input Representation}

\begin{figure}[htbp]
    \centering
        \begin{minipage}{0.95\linewidth}
            \centering
            \includegraphics[width=1\linewidth]{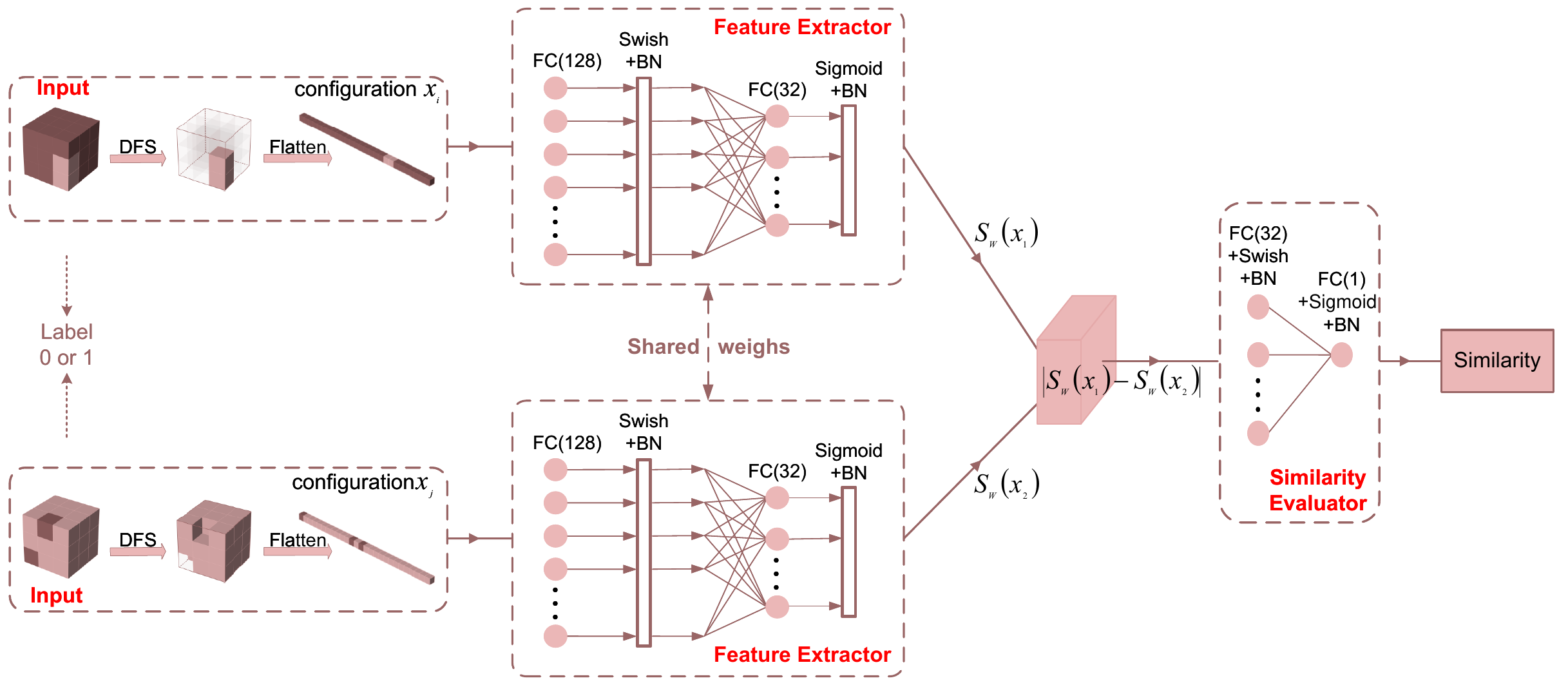}
        \end{minipage}%
    \caption{Schematic Diagram of the SNN Architecture.}
    \label{SNN}
\end{figure}

To provide physically meaningful input to the SNN, we preprocess each raw percolation configuration by identifying its largest connected cluster. This is achieved using a classical graph traversal algorithm known as depth-first search (DFS), which systematically explores neighboring occupied sites to construct connected components\cite{tarjan1972depth}. Among all clusters in a configuration, we retain only the largest one, as it best reflects the emergence of long-range connectivity—a hallmark of the percolation transition. The resulting largest cluster is then flattened into a one-dimensional vector for input into the Siamese network, as shown in the “Input” section on the far left of Figure~\ref{SNN}.

This preprocessing step is critical. As demonstrated in our previous work on two-dimensional systems~\cite{xu2025identifying}, machine learning models trained on raw configurations tend to capture the global occupation density rather than the critical behavior. This occurs because such models are inherently sensitive to the density of active sites, yielding a linear response that obscures the phase transition signal. By extracting the largest cluster, we filter out the linear contribution of isolated clusters and isolate the non-trivial density evolution characteristic of the order parameter $P_{\infty}$. We provide a quantitative validation of this design choice for 3D systems in Section \ref{sec43}.

It is worth noting that while we employ DFS for this extraction, classical cluster-labeling algorithms such as the Hoshen-Kopelman and Newman-Ziff algorithms are computationally more efficient for standard lattices. However, in our workflow, cluster extraction serves solely as a one-time preprocessing step. The choice of traversal algorithm does not influence the SNN's training dynamics or prediction accuracy, as the resulting largest cluster data remains identical regardless of the extraction method.

\vspace{0.3em}
{\bf 2) SNN Architecture for Percolation Model Analysis}

As illustrated in Figure \ref{SNN}, the SNN architecture designed for identifying critical points in percolation models comprises two parallel fully connected neural networks (FCNNs) as "feature extractors", followed by a "similarity evaluator". Importantly, the fully connected design does not impose constraints on the dimensionality or topology of the input, enabling seamless deployment of the same architecture across both 2D and 3D systems.

The fundamental principle of our framework is to ascertain the critical point of a phase transition by comparing the similarity between different percolation configurations using SNN. Firstly, each configuration is subjected to preprocessing via DFS to extract its largest connected cluster. The extracted cluster is then flattened into a one-dimensional vector for input. These preprocessed vectors are subsequently fed into the aforementioned pair of identical, weight-sharing "feature extractors" (i.e., FCNNs), as depicted in the middle section of Figure~\ref{SNN}.

Each constituent branch is comprised of two fully connected layers, with 128 and 32 neurons respectively, each of which is followed by a batch normalization layer and an activation function. The initial activation function employed is a Swish function, the purpose of which is to enhance nonlinearity, and the subsequent activation function is a sigmoid function, the purpose of which is to constrain the output to the range [0, 1]. The architecture of each branch is as follows:
\begin{equation}
\left\{ \text{Fl} - \text{FC}(128) - \text{B} - \text{Ac}(\text{Swish}) - \text{FC}(32) - \text{B} - \text{Ac}(\text{Sigmoid}) \right\}.
\end{equation}
\noindent{Here, Fl denotes the flatten operation, FC(n) denotes a fully connected layer with $n$ neurons, B denotes the batch normalization (BN) layer, and Ac(name) represents the activation function. Each input configuration is thus encoded into a 32-dimensional embedding vector, denoted \( S_\text{repr}(x) \), which captures the structural characteristics of the largest cluster.}

To evaluate the similarity between a pair of configurations \( (x_1, x_2) \), we compute the element-wise absolute difference of their embeddings \( |S_\text{repr}(x_1) - S_\text{repr}(x_2)| \), and feed it into a “similarity evaluator” network, as shown on the far right of Figure~\ref{SNN}. This subnetwork consists of:
\begin{equation}
\left\{ \text{FC}(32) - \text{B} - \text{Ac}(\text{Swish}) - \text{FC}(1) - \text{B} - \text{Ac}(\text{Sigmoid}) \right\},
\end{equation}
\noindent which compresses the input into a scalar similarity score in the range [0, 1]. A score close to 1 indicates that the two configurations are likely from the same phase, while a score near 0 suggests they likely belong to different phases.

\vspace{0.3em}
{\bf 3)Similarity Learning and Optimization}

Given an input pair of percolation configurations (\(x_1\), \(x_2\)) with a binary label \(y \in \{0,1\}\), where \(y=1\) denotes that the two configurations belong to the same phase (referred to as a “positive pair”), and \(y=0\) means they belong to different phases (a “negative pair”). These labels are used to supervise the similarity learning process by serving as targets in the loss function. The SNN learns a shared transformation, parameterized by a weight matrix \(W\), to map both inputs into a common latent representation space. A simplified form of this mapping can be expressed as:

\begin{equation}
S_W(x) = \sigma(Wx + b),
\end{equation}

\noindent where $\sigma$ is the activation function (e.g., Swish or Sigmoid), and $b$ is the bias term. 
Instead of computing a predefined scalar distance (such as Euclidean or Cosine distance), we compute the element-wise absolute difference vector $\Delta(x_1, x_2)$ between the latent representations:

\begin{equation}
\Delta(x_1, x_2) = |S_{W}(x_1) - S_{W}(x_2)| \in \mathbb{R}^{n},
    \label{eq6}
\end{equation}
where $n$ is the dimension of the embedding space (here $n=32$), and $|\cdot|$ denotes the element-wise absolute value operation. This difference vector $\Delta(x_1, x_2)$ preserves the distinct variations along each feature dimension. It is subsequently fed into the similarity evaluation network (see the far-right section of Figure~\ref{SNN}), which maps this vector to a scalar similarity score $s(x_1, x_2) \in [0, 1]$ through learnable weights.

This design choice constitutes a "learnable metric" approach. Unlike fixed metrics (e.g., Euclidean, Cosine, or Wasserstein distances) or topological metrics (e.g., Betti numbers) which impose a rigid, predefined notion of similarity, our approach allows the model to adaptively learn the importance of different feature dimensions. By feeding the full difference vector into a subsequent neural network, the system can capture complex, non-linear similarity patterns specific to percolation clusters that might be overlooked by standard geometric metrics. This end-to-end learning strategy enables the SNN to discover the optimal similarity function directly from the data, rather than relying on manual metric selection.

\vspace{0.3em}
{\bf 4)Training and Loss Function}

The SNN is trained using the binary cross-entropy loss function, defined as:

\begin{equation}
L = -\frac{1}{N} \sum_{i=1}^{N} \left[ y(x_i,x_j) \log(s(x_i,x_j)) + (1 - y(x_i,x_j)) \log(1 - s(x_i,x_j)) \right],
\label{eq7}
\end{equation}

\noindent where $N$ is the batch size, $y(x_i, x_j)$ is the true label assigned to the $i$-th input pair, indicating whether the two configurations belong to the same phase ($y=1$) or not ($y=0$), and $s(x_i, x_j)$ is the predicted similarity. During backpropagation, the gradient of the loss with respect to the weights $W$ is computed as:

\begin{equation}
\frac{\partial L}{\partial W} = \frac{1}{N} \sum_{i=1}^{N} \left[ \left( s(x_i,x_j) - y(x_i,x_j) \right) \cdot \frac{\partial s(x_i,x_j)}{\partial W} \right],
\end{equation}
\noindent enabling the optimization of $W$ to minimize the loss and improve model performance.

\section{Results}

\subsection{Monte Carlo Results}
\label{sec41}

\begin{figure}[htbp]
    \centering
    \subfigure[]{
            \begin{minipage}[b]{0.45\textwidth}
                \centering  
                \includegraphics[width=0.9\textwidth]{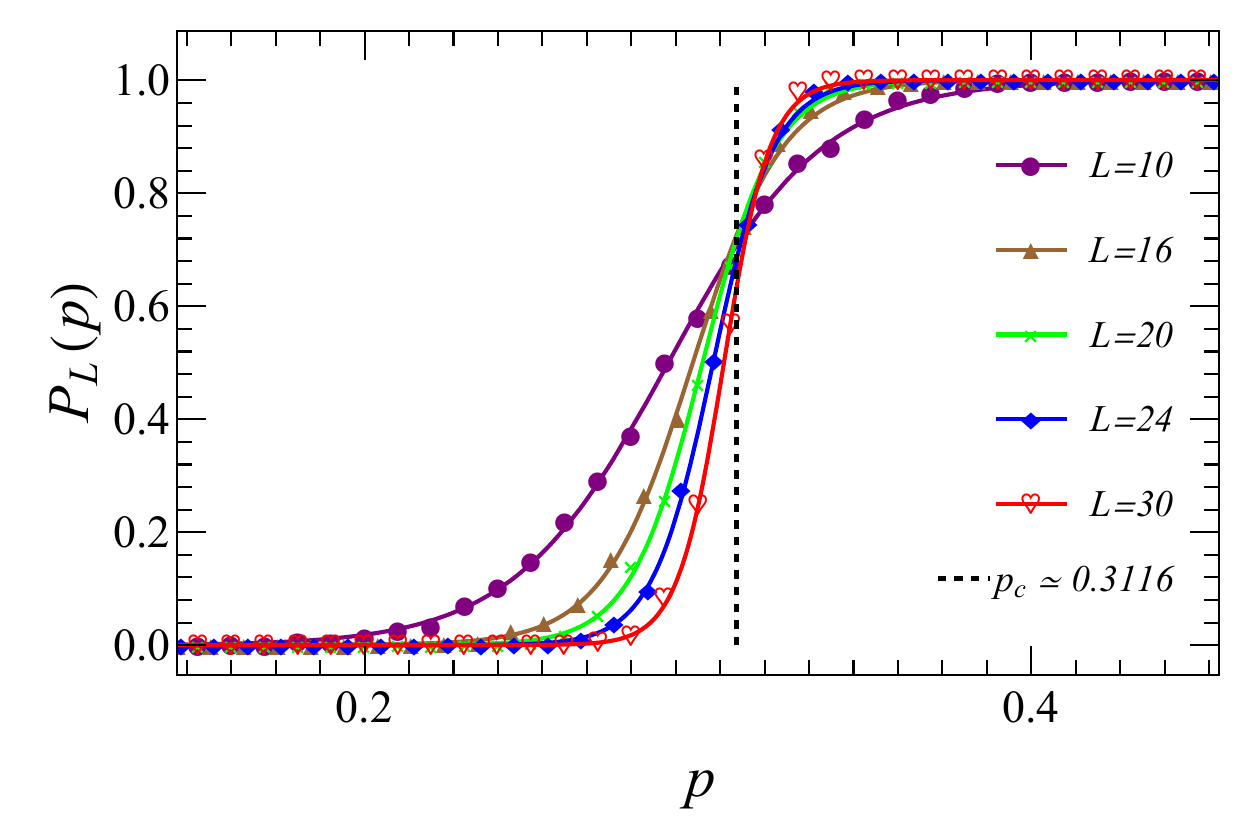} 
            \end{minipage}%
        }
    \subfigure[]{
        \begin{minipage}[b]{0.45\textwidth}
            \centering
            \includegraphics[width=0.9\textwidth]{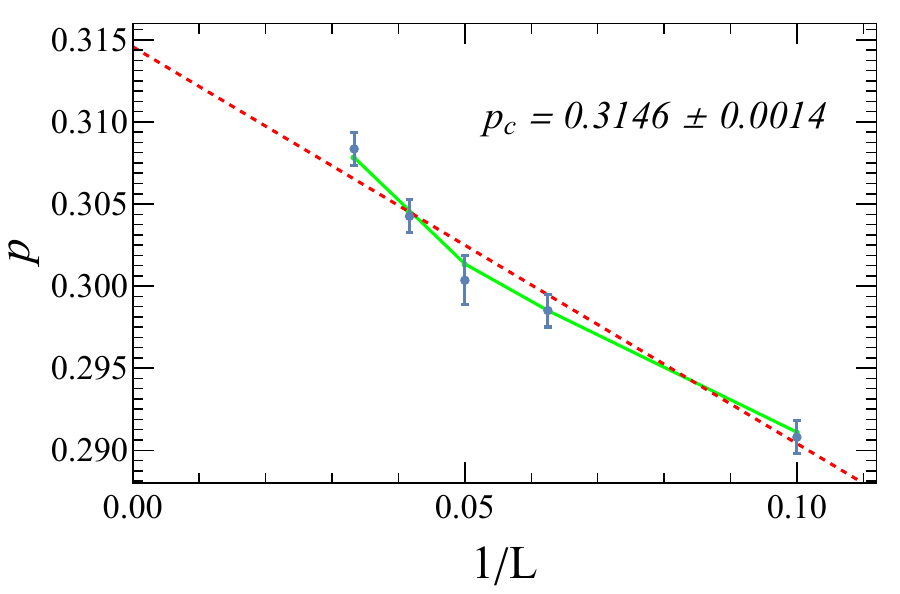}
        \end{minipage}
    }
    \caption{
    Monte Carlo simulation results for the 3D site percolation model. Panel (a) shows the percolation probability $P_L(p)$ as a function of the site occupation probability $p$ for system sizes $L=10,16,20,24,$ and $30$. Here $P_L(p)$ denotes the percolation probability, defined as the fraction of realizations that contain at least one spanning cluster connecting opposite sides of the system. Each curve was fitted by Eq.~\eqref{tanh}.
    All sigmoid fits achieve a goodness of fit exceeding 99.9\%.
    The fitted critical thresholds are
    $p_c(10)=0.2911$, $p_c(16)=0.2985$, $p_c(20)=0.3014$, $p_c(24)=0.3046$, and $p_c(30)=0.3078$.
    The dashed vertical line marks the theoretical threshold.
    Panel (b) presents the extrapolation of $p_c(L)$ to the thermodynamic limit using FSS, yielding $p_c^{(\infty)} = 0.3146 \pm 0.0014$.}
    \label{3dsiteMC}
\end{figure}

\begin{figure}[htbp]
    \centering
    \subfigure[]{
        \begin{minipage}[b]{0.45\textwidth}
            \includegraphics[width=0.9\textwidth]{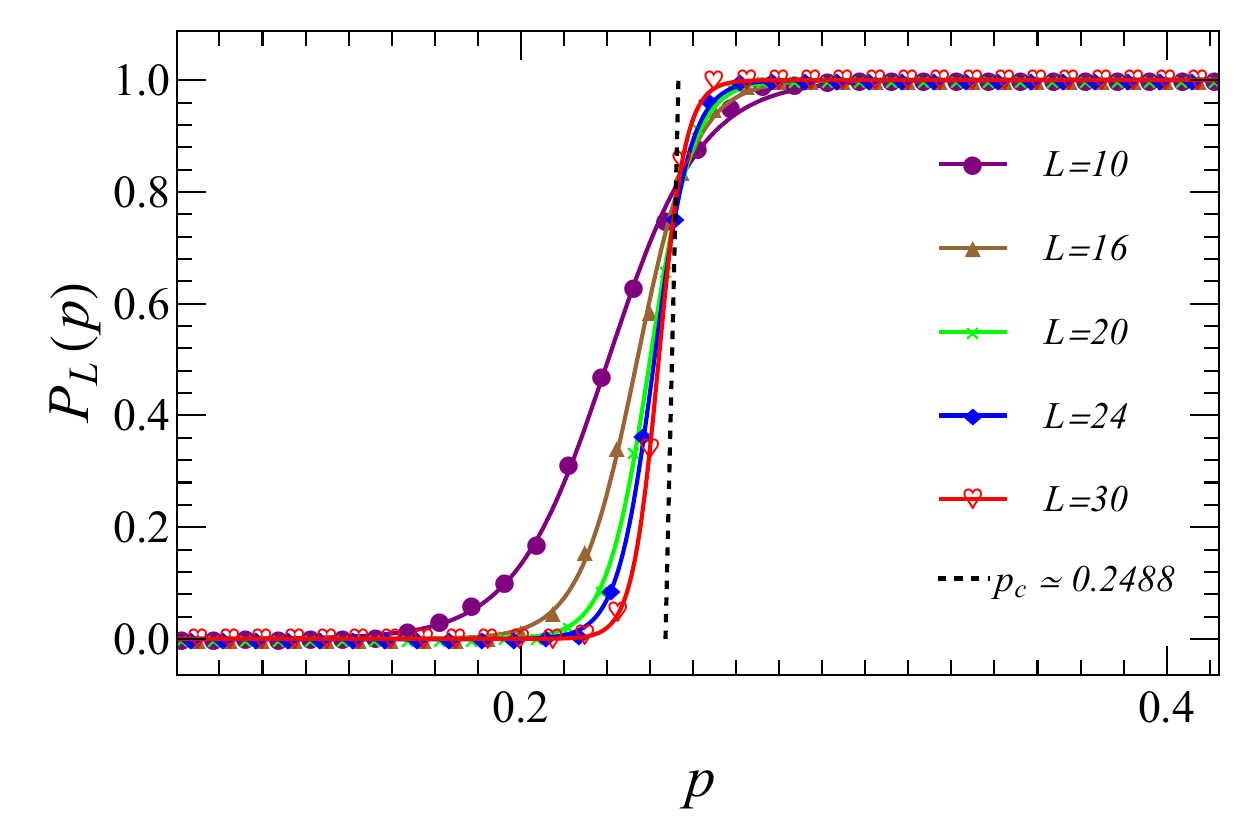}
        \end{minipage}
    }
    \subfigure[]{
        \begin{minipage}[b]{0.45\textwidth}
            \includegraphics[width=0.9\textwidth]{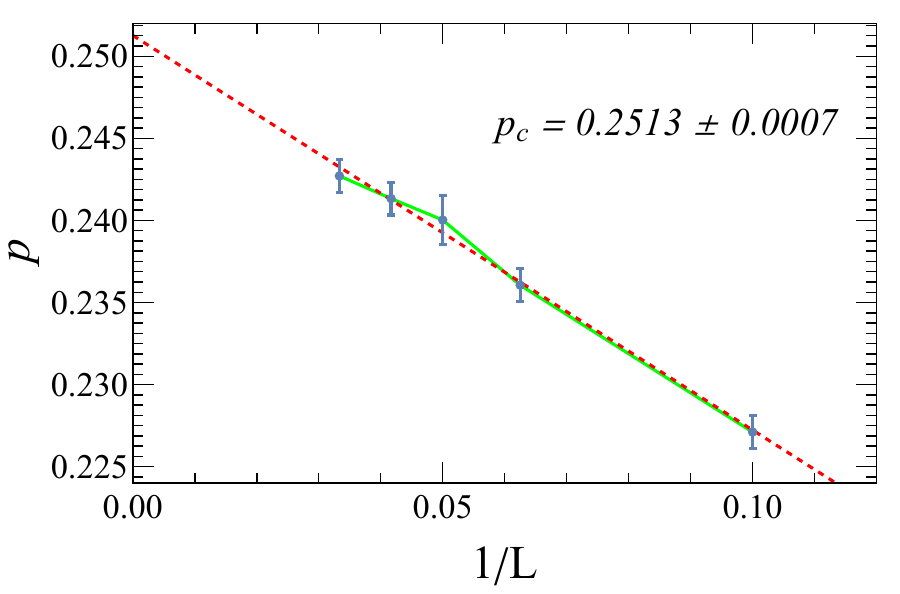}
        \end{minipage}
    }
    \caption{
    Monte Carlo simulation results for the 3D bond percolation model.Panel (a) shows the percolation probability $P_L(p)$ as a function of the bond occupation probability $p$ for system sizes $L=10,16,20,24,$ and $30$. Here $P_L(p)$ denotes the percolation probability, defined as the fraction of realizations that contain at least one spanning cluster connecting opposite sides of the system. Each curve was fitted by Eq.~\eqref{tanh}.
    All sigmoid fits achieve a goodness of fit exceeding 99.9\%.
    The fitted critical thresholds are
    $p_c(10)=0.2271$, $p_c(16)=0.2361$, $p_c(20)=0.2400$, $p_c(24)=0.2413$, and $p_c(30)=0.2417$.
    The dashed vertical line marks the theoretical threshold.
    Panel (b) presents the extrapolation of $p_c(L)$ to the thermodynamic limit using FSS, yielding $p_c^{(\infty)} = 0.2513 \pm 0.0007$.}
    \label{3dbondMC}
\end{figure}

Prior to conducting ML analysis, we performed Monte Carlo simulations on 3D site and bond percolation models defined on cubic lattices to validate the reliability of the generated configurations. Simulations were carried out for system sizes \(L = 10, 16, 20, 24,\) and \(30\), generating 1000 independent samples per occupation probability \(p\) to reduce statistical uncertainties. 
The critical thresholds \(p_c(L)\) for each system size were determined by fitting the percolation probability \(P_L(p)\), as illustrated in Figs.~\ref{3dsiteMC}(a) and~\ref{3dbondMC}(a), which is defined as the fraction of independent Monte Carlo realizations that contain at least one spanning cluster under open boundary conditions. This quantity exhibits a smooth transition from 0 to 1 near the finite-size critical point \(p_c(L)\), and the finite-size critical points \(p_c(L)\) obtained from these fits were further analyzed using FSS to estimate the thermodynamic limit \(p_c\).


To quantitatively locate \(p_c(L)\), the curves of \(P_L(p)\) were fitted by a sigmoidal function of the form
\begin{equation}
    P_L(p) = a\,\tanh[b(p - p_c)] + c,
    \label{tanh}
\end{equation}
where \(a\) and \(c\) control the vertical scaling and offset of the curve, and \(b\) characterizes the sharpness of the transition. The fitted midpoint \(p_c\) corresponds to the finite-size critical threshold. Nonlinear least-squares fitting was performed using Mathematica's \textit{NonlinearModelFit} function. For each system size, initial parameter guesses were provided to ensure convergence (e.g., $\{a \approx -0.5, b \approx 10, c \approx 0.5, p_c \approx 0.3\}$ for site percolation). The goodness of fit was quantified by the coefficient of determination $R^2 = 1 - \text{RSS}/\text{TSS}$, where RSS is the residual sum of squares and TSS is the total sum of squares. This empirical model provides a smooth approximation of the finite-size crossover from the non-percolating to the percolating regime, and all fitted curves achieved a goodness of fit exceeding 99.9\%.

The obtained size-dependent thresholds $p_c(L)$ were then extrapolated to the thermodynamic limit following the standard finite-size scaling ansatz:
\begin{equation}
    p_c(L) = p_c^\infty + A/L, 
\end{equation}
where $p_c^\infty$ denotes the critical threshold as $L \to \infty$. A weighted linear regression yielded the asymptotic value $p_c^\infty$ (intercept) and the system-dependent constant $A$ (slope). The uncertainty in $p_c^\infty$ (e.g., $0.3146 \pm 0.0014$ for site percolation) is reported as the standard error of the intercept. The validity of this linear extrapolation is confirmed by the distinct linear trends observed in Figures~\ref{3dsiteMC}(b) and~\ref{3dbondMC}(b).
These results align well with theoretical predictions (\(p_c = 0.31160768(15)\) and \(p_c = 0.248811 85(10)\), respectively~\cite{xu2014simultaneous}), providing confidence in the validity of the simulation framework for downstream ML analysis. Based on these verified datasets, we proceeded with further analysis using the SNN approach.

\subsection{Input Data and Preprocessing for SNN}
\label{sec42}

\vspace{0.5em}
To facilitate the computational analysis and enable SNN-based learning, we generated binary tensor representations for the 3D site and bond percolation models:

Site percolation: A binary tensor of shape $L \times L \times L$, where each element is either 1 (occupied site) or 0 (vacant site), representing the occupancy status of each site in a 3D cubic lattice.

Bond percolation: Bond configurations involve connections between adjacent sites, resulting in an irregular structure that cannot be directly mapped to a standard $L \times L \times L$ tensor. To unify the input format for the neural network, we represent bond configurations using an expanded cubic tensor of size $(2L+1)^3$. This tensor employs a ternary encoding scheme: `1' represents an occupied bond, `0' represents a physically existing but unoccupied bond, and `2' serves as a padding value for non-physical spatial gaps.

The introduction of the padding value `2' is a specific design choice to avoid semantic ambiguity. Since `0' already carries the physical meaning of a "blocked path" (an existing bond that is not occupied), using a binary encoding (where `0' also represents padding) would prevent the model from distinguishing between valid bond positions and void spaces. By designating `2' as a unique identifier for non-physical positions, we ensure that the topological structure is correctly preserved. During the machine learning process, a Boolean mask is applied to strictly ignore these padding values, ensuring they do not contribute to feature extraction or model training.

For each percolation probability $p$, 1000 independent configurations were generated and stored individually, resulting in 101 data points spanning the range $p \in [0,1]$. The largest cluster from each configuration was extracted using the DFS algorithm. These extracted configurations were subsequently used as input pairs for training and evaluating the SNN. 

\vspace{1em}
\noindent {\bf Labeling Strategy for Semi-Supervised Learning}

This study adopts a semi-supervised learning framework for training the SNN, where only a small subset of the dataset is labeled. To minimize manual intervention and avoid introducing physical bias near the critical point, labeled samples were selected exclusively from the regions far from criticality, specifically the intervals \( [0, 0.1] \cup [0.9, 1] \). This strategy resulted in only 22 labeled probability points across multiple system sizes. 

Unlike traditional supervised learning, which assigns labels to individual samples, the SNN adopts a pairwise labeling strategy. Labels are assigned to configuration pairs based on their relative similarity or dissimilarity.
Specifically, two probabilities $p_i$ and $p_j$ are randomly sampled from within the intervals $[0,0.1]$ and $[0.9,1]$, respectively.
The corresponding configurations $x_i$ and $x_j$ are then paired for similarity evaluation. The pair is labeled as:
\begin{itemize}
  \item Positive: if both $x_i$ and $x_j$ are sampled from the same probability interval—either from the low-probability regime $[0, 0.1]$ or the high-probability regime $[0.9, 1]$, they are labeled as a positive pair.

  \item Negative: if $x_i$ and $x_j$ are sampled from different intervals (i.e., one from $[0, 0.1]$ and the other from $[0.9, 1]$), they are labeled as a negative pair.
\end{itemize}
To prevent sampling bias, the pool of 1000 configurations at each $p$ value was independently shuffled prior to pair generation.

\vspace{1em}
\noindent {\bf Training Process}

1. The selected configurations $x_i$ and $x_j$ are input to the two branches of the SNN feature extractor, which maps them to latent representations $S_W(x_i)$ and $S_W(x_j)$.

2. The element-wise absolute difference vector is computed (Eq.~\ref{eq6}), capturing the specific variations between the two latent embeddings across all dimensions.

3. This difference vector is then fed into the similarity evaluator network, which learns to map these variations to a scalar similarity score $s(x_i, x_j) \in [0, 1]$ via a Sigmoid activation layer.

4. Model parameters (including both the feature extractor and the similarity evaluator) are optimized by minimizing the binary cross-entropy loss function (Eq.~\ref{eq7}) via backpropagation.

\vspace{1em}
\noindent {\bf Testing Phase}

1. Randomly select a reference probability \( p \in [0,1] \), and use its corresponding configuration as a reference. This reference is then paired with configurations sampled at various other probabilities \( p_i \in [0,1] \), and each pair is fed into the SNN. This reference probability \( p \) is defined as the anchor.

2. The SNN computes the similarity $s(x_p, x_i)$ between the anchor configuration $x_p$ and the test configuration $x_i$.

3. By plotting the similarity $s(x_p, x_i)$ as a function of $p_i$, the critical threshold $p_c$ is identified as the intersection point between the average similarity curves of positive and negative sample pairs.

\subsection{SNN Results}
\label{sec43}

\vspace{0.3em}
\textbf{1) Raw input}

Before analyzing the results from the DFS-extracted features, we briefly address the necessity of this preprocessing step. To respond to the need for a quantitative assessment, we tested the SNN with raw configuration inputs (binary tensors) on 3D site percolation lattices ($L=16, 20$). 

\begin{figure}[htbp]
    \centering
        \begin{minipage}{1\linewidth}
            \centering
            \includegraphics[width=0.35\textwidth]{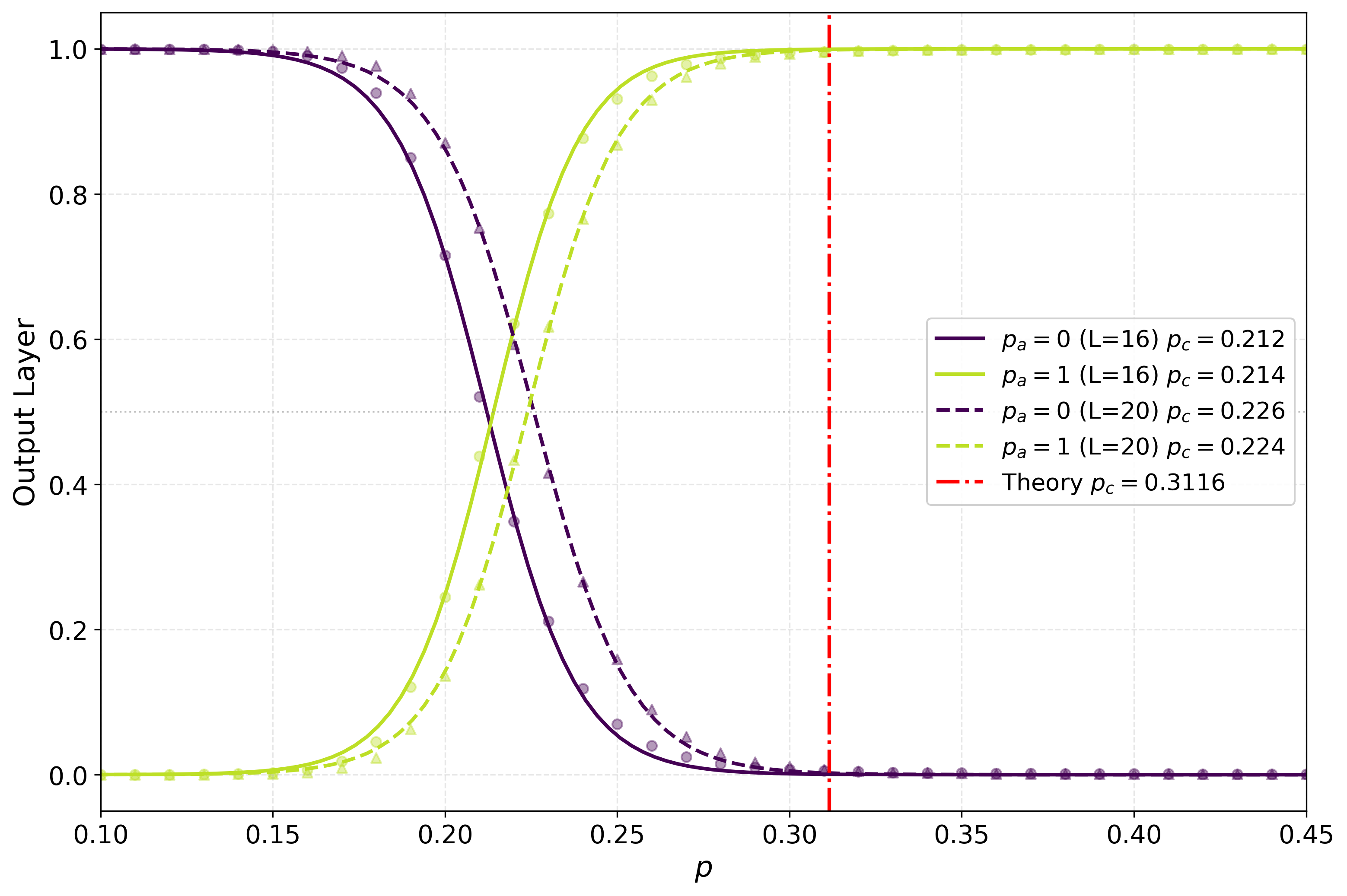}
            \caption{SNN learning results on raw configuration inputs for 3D site percolation with system sizes $L=16$ and $L=20$. The model predicts a crossover at $p \approx 0.22$, far from the theoretical critical point $p_c \approx 0.3116$ (red dashed line). This failure validates the necessity of the DFS-based largest cluster extraction used in our methodology.}
            \label{raw_input}
        \end{minipage}
\end{figure}
As shown in Figure~\ref{raw_input}, our experiments revealed that models trained on raw inputs failed to locate the correct critical threshold. Specifically, the predicted transition point shifted to $p \approx 0.22$, deviating significantly from the theoretical value $p_c = 0.311 607 68(15)$. This quantitative discrepancy confirms that, consistent with our previous findings in 2D systems, neural networks fed with raw configurations tend to overfit to the local site density rather than capturing the long-range topological connectivity. Consequently, DFS-based extraction is not merely a design choice but a prerequisite for correctly identifying the phase transition in 3D percolation.

\vspace{0.3em}
\textbf{2)  Largest cluster input}

\begin{figure}[htbp]
    \vspace{-0.7em}
	\centering
	\subfigure[$p_a=0.47$]{
		\begin{minipage}[b]{0.27\textwidth}
			\includegraphics[width=0.95\textwidth]{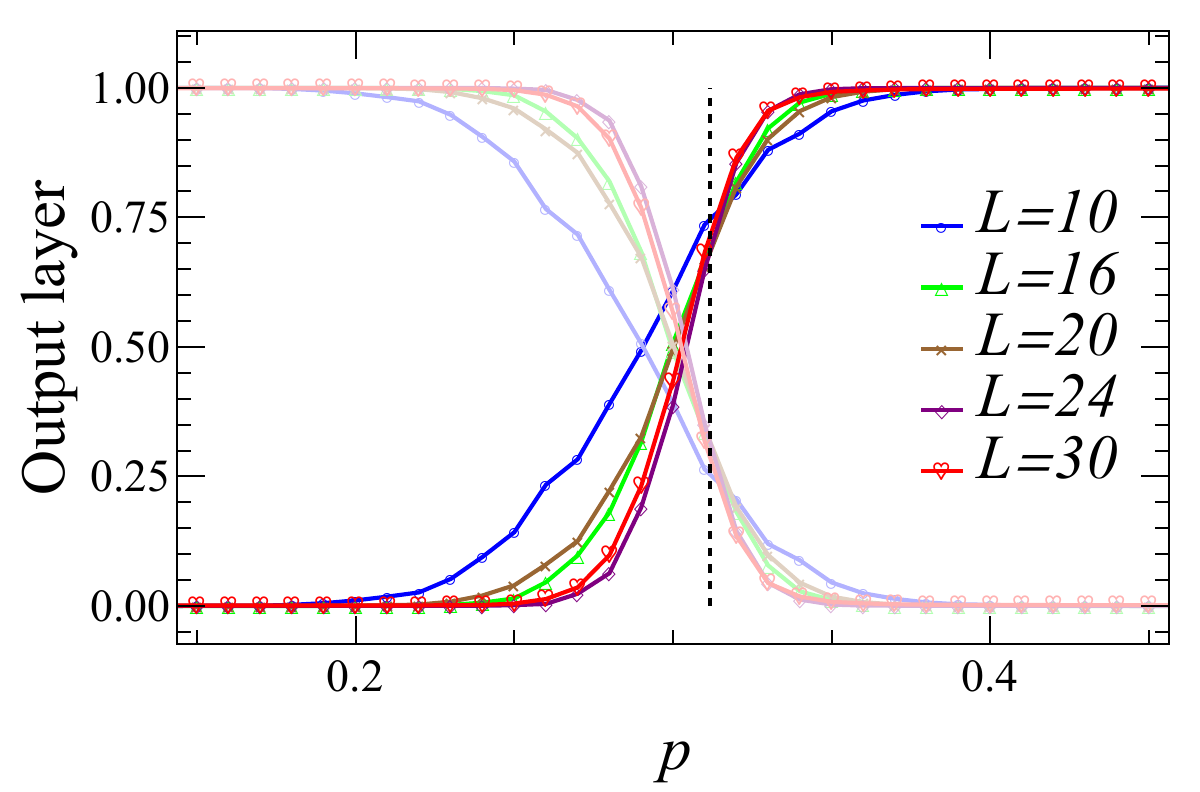} 
		\end{minipage}
	}
	\subfigure[$p_a=0.47$]{
		\begin{minipage}[b]{0.27\textwidth}
			\includegraphics[width=0.95\textwidth]{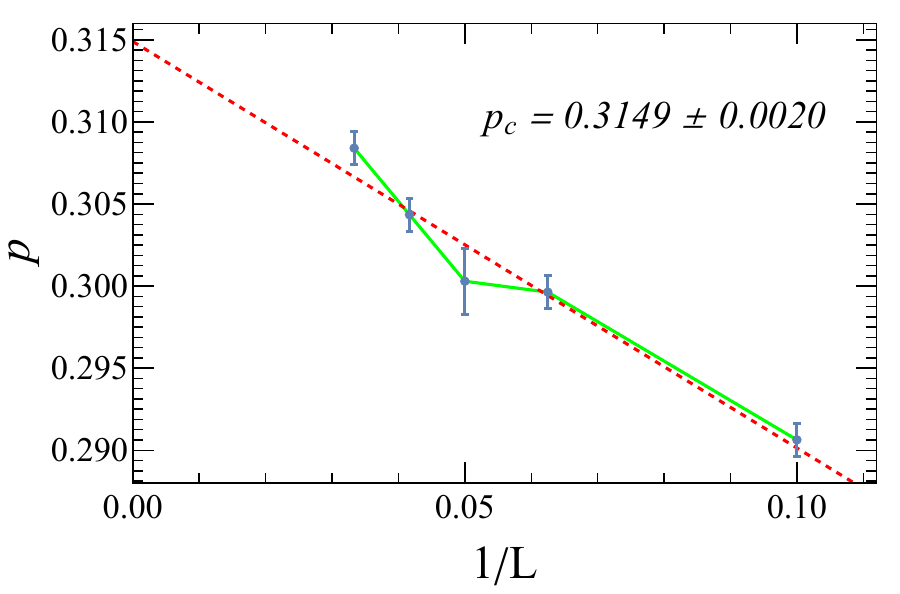}
		\end{minipage}
	}
		\subfigure[$p_a=0.31$]{
		\begin{minipage}[b]{0.27\textwidth}
			\includegraphics[width=0.95\textwidth]{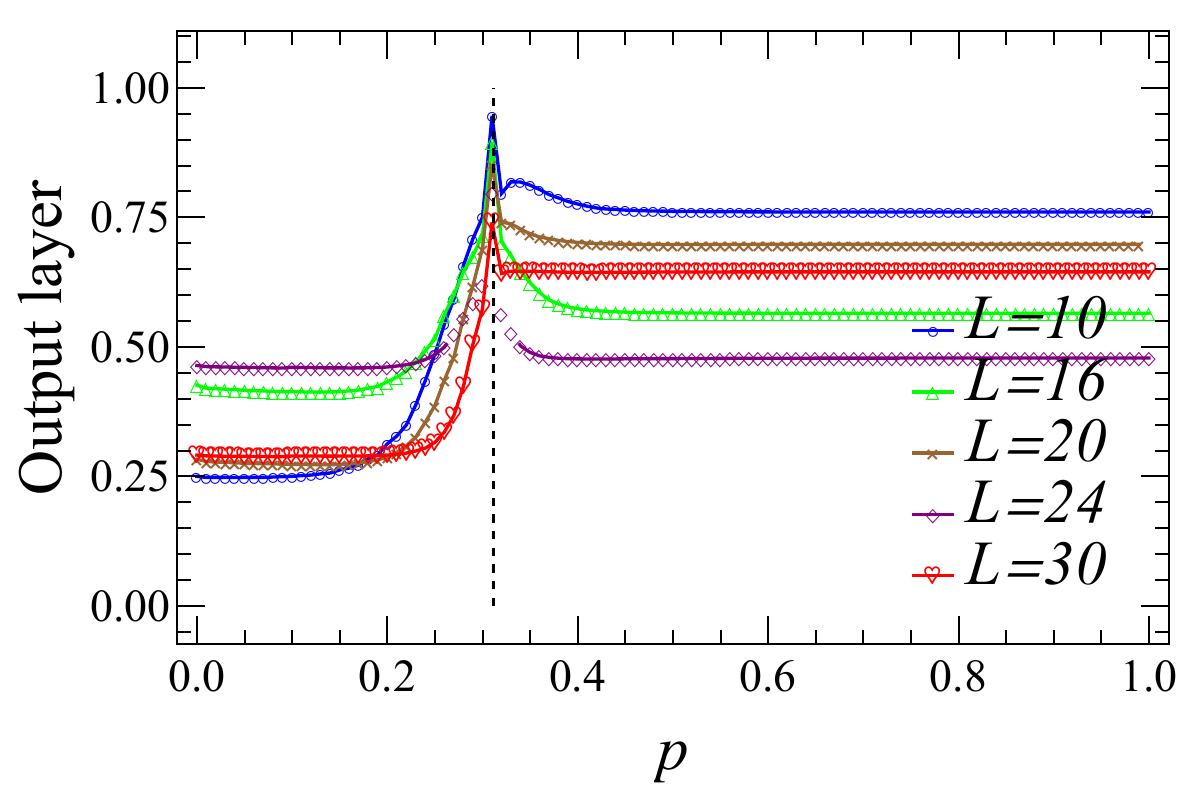}
		\end{minipage}
	}
	\\
		\subfigure[$p_a=0.12$]{
		\begin{minipage}[b]{0.27\textwidth}
			\includegraphics[width=0.95\textwidth]{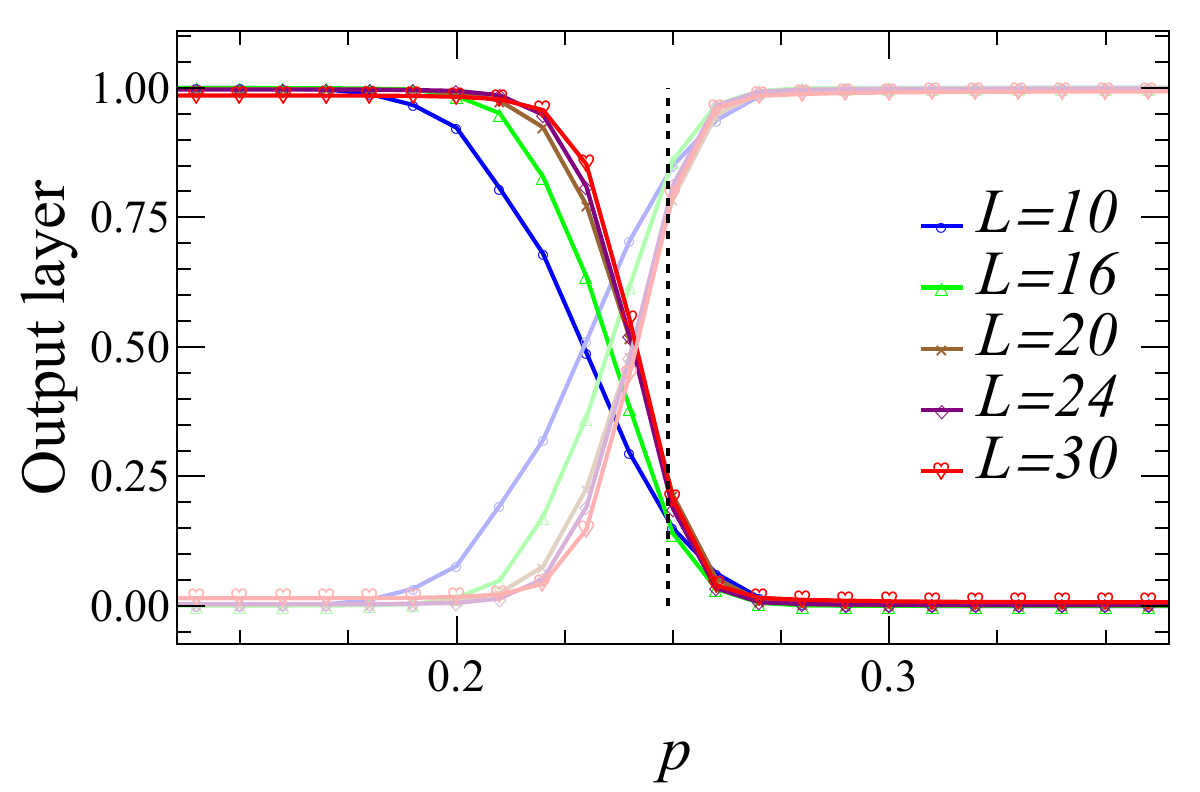} 
		\end{minipage}
	}
	\subfigure[$p_a=0.12$]{
		\begin{minipage}[b]{0.27\textwidth}
			\includegraphics[width=0.95\textwidth]{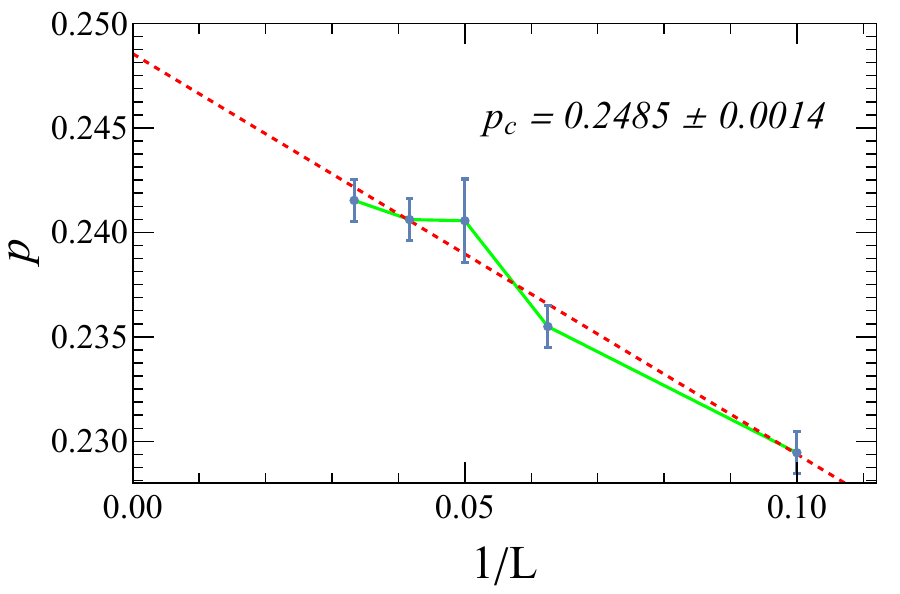}
		\end{minipage}
	}
	\subfigure[$p_a=0.25$]{
		\begin{minipage}[b]{0.27\textwidth}
			\includegraphics[width=0.95\textwidth]{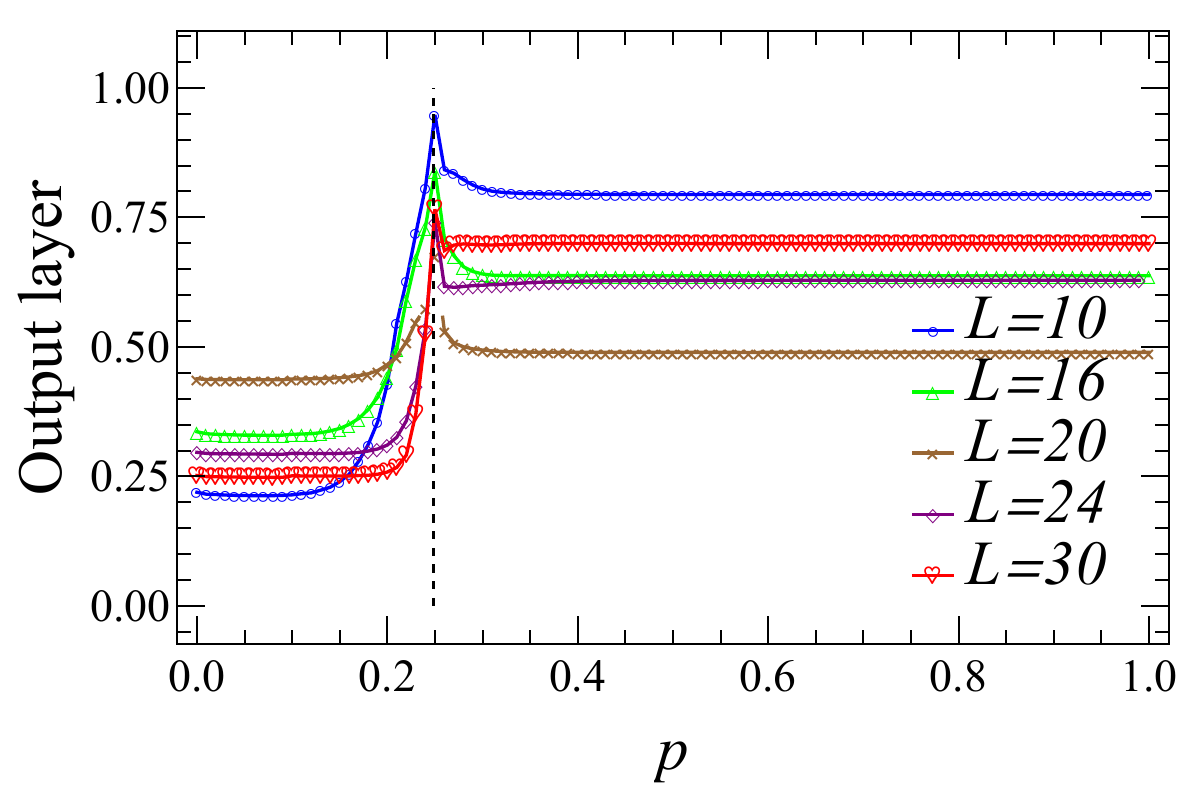}
		\end{minipage}
	}
	\vspace{-0.7em}
    \caption{The first row shows results for the three-dimensional site percolation model:
    (a) Similarity (dark) and dissimilarity (light) curves for various system sizes at anchor \( p_a = 0.47 \);
    (b) FSS extrapolation for anchor \( p_a = 0.47 \);(c) Output layer responses for various system sizes at anchor \( p_a = 0.31 \), where the similarity curve peaks near the critical threshold \( p_c \approx 0.31 \).
    The second row shows results for the three-dimensional bond percolation model:
    (d) Similarity (dark) and dissimilarity (light) curves for various system sizes at anchor \( p_a = 0.12 \);
    (e) FSS extrapolation for anchor \( p_a = 0.12 \); (f) Output layer responses for various system sizes at anchor \( p_a = 0.25 \), where the similarity curve peaks near \( p_c \approx 0.25 \).}
    \label{3dsite}
\end{figure}

In the three-dimensional site percolation model, an anchor probability of $p_a = 0.47$ was selected as the reference for similarity evaluation.The similarity computed by the SNN for various system sizes is shown in Figure~\ref{3dsite}(a). 
The results indicate that the similarity curve remains relatively stable in the intervals $[0, 0.2]$ and $[0.4, 1]$, but exhibits a sharp rise within $[0.2, 0.4]$. This interval is identified as the phase-transition region. The critical threshold in the thermodynamic limit, denoted as $p_c^\infty$, was estimated using FSS, as shown in Figure~\ref{3dsite}(b). Similarly, for the three-dimensional bond percolation model with an anchor probability of $p_a = 0.12$, the similarity curves and the estimated threshold $p_c^\infty$ are presented in Figures~\ref{3dsite}(d) and \ref{3dsite}(e), respectively. For both site and bond percolation models, 
our estimated critical thresholds are consistent with standard theoretical values within statistical uncertainty at the percent level. While this precision does not match the state-of-the-art Monte Carlo results (which reach up to six decimal places), it confirms that the SNN successfully captures the correct critical physics from limited samples. This supports the validity of the proposed framework as a methodological proof-of-concept.

To assess the impact of anchor selection on prediction accuracy, multiple anchors were chosen from non-critical (non-phase-transition) regions for evaluation (see Tables~\ref{3dsitepc} and \ref{3dbondpc}). 
\begin{table}[htbp]
\caption{\hl{Critical thresholds for 3D site percolation by anchor.}}
\label{3dsitepc}
\rmfamily
\begin{tabular*}{\tblwidth}{@{}LLLLLLL@{}}
\toprule
~ & $L=10$ & $L=16$ & $L=20$ & $L=24$ & $L=30$ & $p^\infty_c$ \\ 
\midrule
$p_a=0$    & 0.2910 & 0.2995 & 0.2985 & 0.3019 & 0.3051 & 0.3103(19) \\
$p_a=0.15$ & 0.2924 & 0.2990 & 0.2982 & 0.3016 & 0.3050 & 0.3090(20) \\
$p_a=0.47$ & 0.2906 & 0.2996 & 0.3003 & 0.3043 & 0.3084 & 0.3149(20) \\
$p_a=1$    & 0.2908 & 0.2998 & 0.3004 & 0.3043 & 0.3084 & 0.3147(20) \\
\bottomrule
\end{tabular*}
\end{table}
\begin{table}
\caption{\hl{Critical thresholds for 3D bond percolation by anchor.}}
\label{3dbondpc}
\rmfamily
\begin{tabular*}{\tblwidth}{@{}LLLLLLL@{}}
\toprule
~ & $L=10$ & $L=16$ & $L=20$ & $L=24$ & $L=30$ & $p^\infty_c$ \\ 
\midrule
$p_a=0$    & 0.2296 & 0.2354 & 0.2405 & 0.2406 & 0.2415 & 0.2484(14) \\
$p_a=0.12$ & 0.2295 & 0.2355 & 0.2405 & 0.2406 & 0.2415 & 0.2485(14) \\
$p_a=0.38$ & 0.2285 & 0.2359 & 0.2427 & 0.2431 & 0.2439 & 0.2530(19) \\
$p_a=1$    & 0.2285 & 0.2362 & 0.2427 & 0.2431 & 0.2438 & 0.2530(18) \\
\bottomrule
\end{tabular*}
\end{table}
The critical thresholds summarized in the tables indicate that variations in anchor selection have only a minor effect on prediction accuracy, demonstrating the robustness of the model. On the other hand, slight fluctuations in the estimated critical threshold across system sizes are observed, likely due to finite-size effects, in accordance with the expectations of FSS theory. 

\vspace{0.3em}
\textbf{3) Sensitivity Analysis of Anchor Selection}

The empirical stability observed in the tables above is fundamentally grounded in the physical concept of correlation length ($\xi$). In the critical region ($p \approx p_c$), $\xi$ diverges, leading to intense fluctuations and high geometric uncertainty. Selecting anchors in this region results in the sharp "spiking" behavior in the similarity curve, as illustrated in Figures~\ref{3dsite}(c) and \ref{3dsite}(f), which introduces instability for precise $p_c$ estimation. In contrast, within non-critical regions, $\xi$ is short, and cluster structures exhibit stable, typical features of the respective phase. Consequently, selecting anchors from these stable regions ensures that the SNN learns the most representative structural features.

\begin{figure}[htbp]
    \centering
        \begin{minipage}{1\linewidth}
            \centering
            \includegraphics[width=0.35\textwidth]{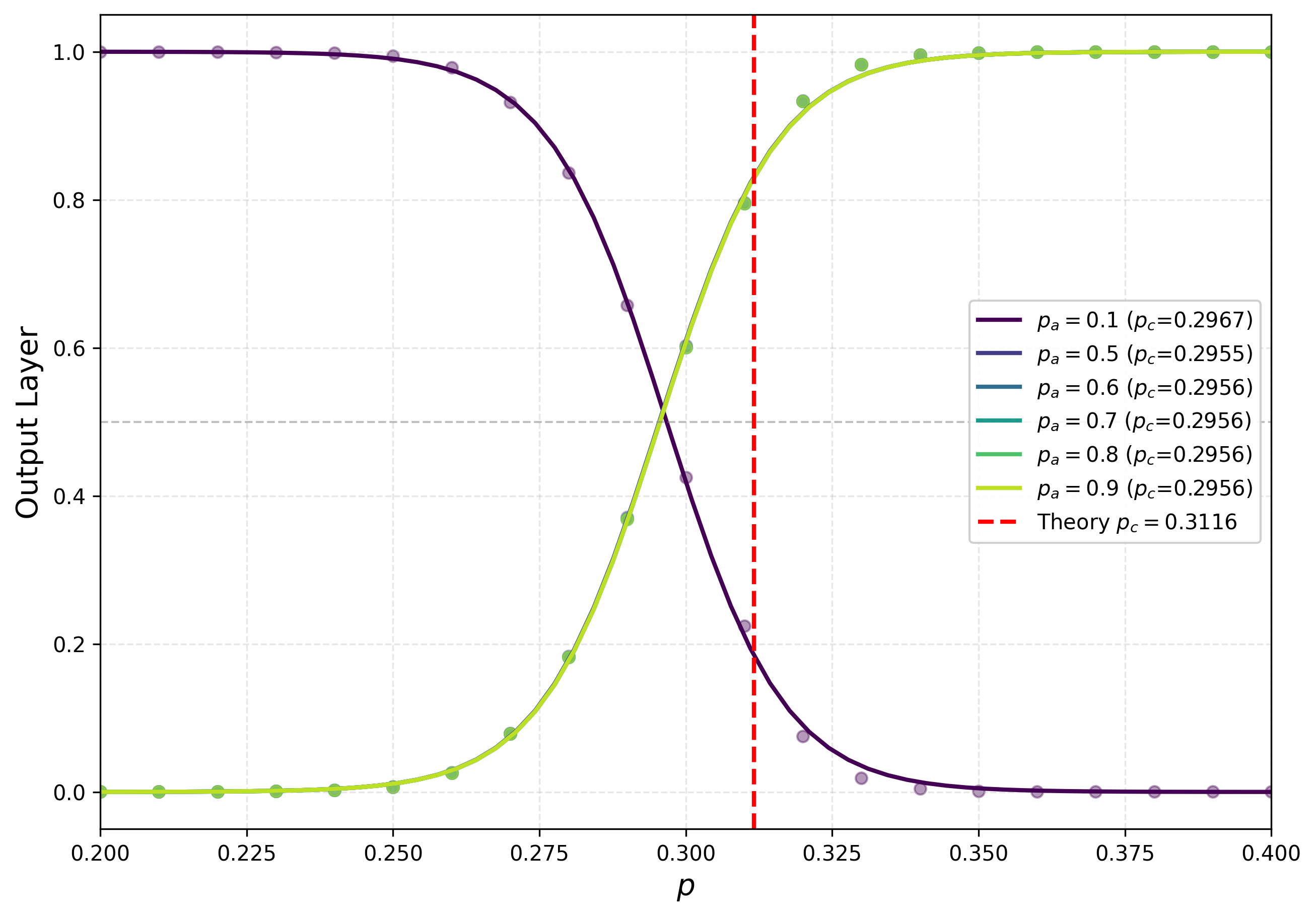}
            
            \caption{Sensitivity analysis of anchor selection. The SNN model trained on SC lattices ($L=20$) is tested using different anchors $p_a \in \{0.1, 0.5, 0.6, 0.7, 0.8, 0.9\}$. All output curves virtually overlap, demonstrating that the method is insensitive to the specific choice of anchors within non-critical regions.}
            \label{fig:anchor_sensitivity}
        \end{minipage}
\end{figure}

To systematically assess this robustness, we conducted a sensitivity analysis on the SC lattice with $L=20$. We selected six anchors from the non-critical regions, $p_a \in \{0.1, 0.5, 0.6, 0.7, 0.8, 0.9\}$. The results are presented in Figure~\ref{fig:anchor_sensitivity}, despite the significant variation in anchor positions, all output response curves exhibit good data collapse, and the predicted critical points fluctuate only within a negligible range $[0.2955, 0.2967]$.

The results highlight a dual behavior in anchor selection. On one hand, the estimation is insensitive to anchor choice within non-critical regions, confirming the method's stability. On the other hand, the emergence of spikes near the critical point provides a reference for identifying the critical fluctuation region.

\vspace{0.3em}
\textbf{4) Data collapse}

\begin{figure}[htbp]
    \centering
        \begin{minipage}{1\linewidth}
            \centering
            \subfigure[$p_a=0$]{
                \begin{minipage}[b]{0.3\linewidth} 
                    \centering
                    \includegraphics[width=0.95\textwidth]{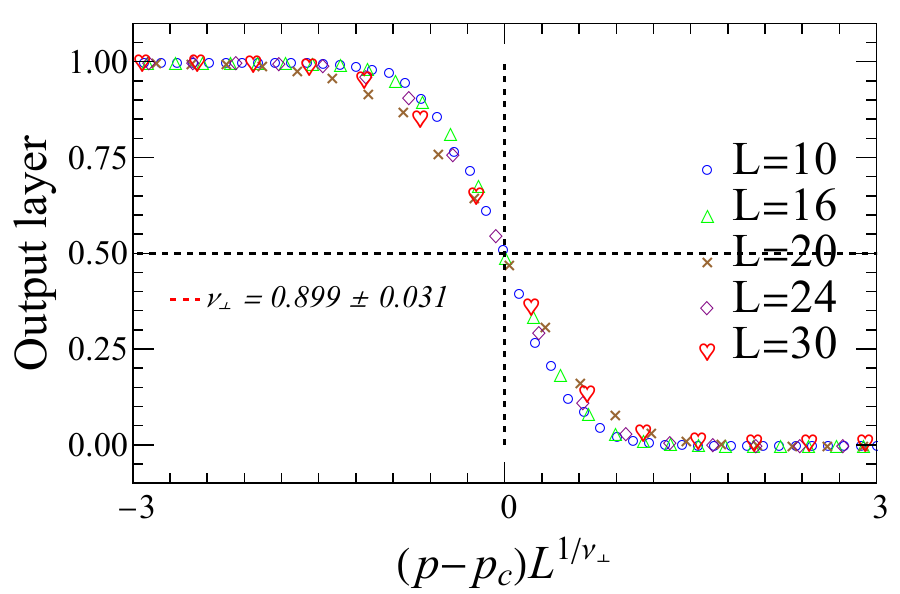}
                \end{minipage}
            }
            \subfigure[$p_a=0.15$]{
                \begin{minipage}[b]{0.3\linewidth}
                    \centering
                    \includegraphics[width=0.95\textwidth]{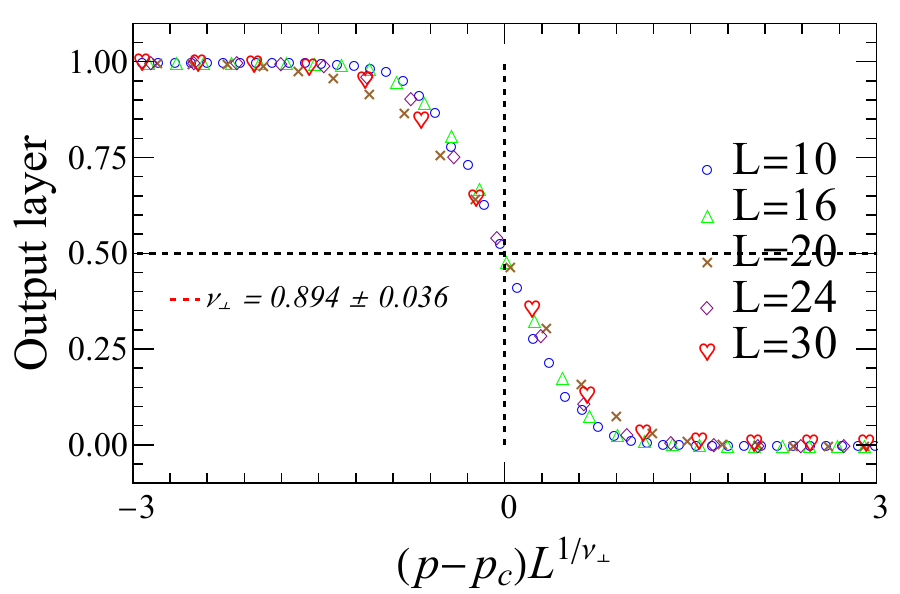}
                \end{minipage}
            }
            \\ 
            \subfigure[$p_a=0.47$]{
                \begin{minipage}[b]{0.3\linewidth}
                    \centering
                    \includegraphics[width=0.95\textwidth]{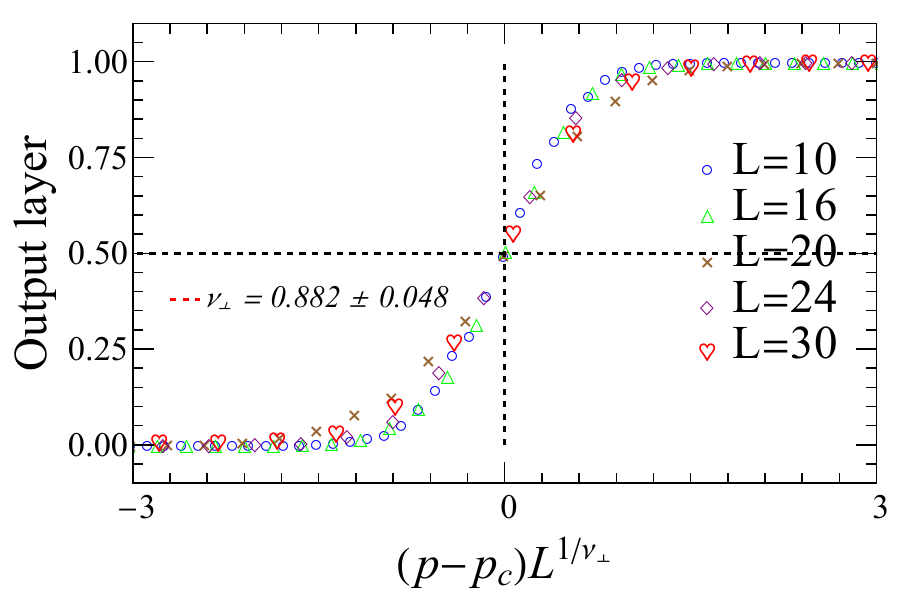}
                \end{minipage}
            }
            \subfigure[$p_a=1$]{
                \begin{minipage}[b]{0.3\linewidth}
                    \centering
                    \includegraphics[width=0.95\textwidth]{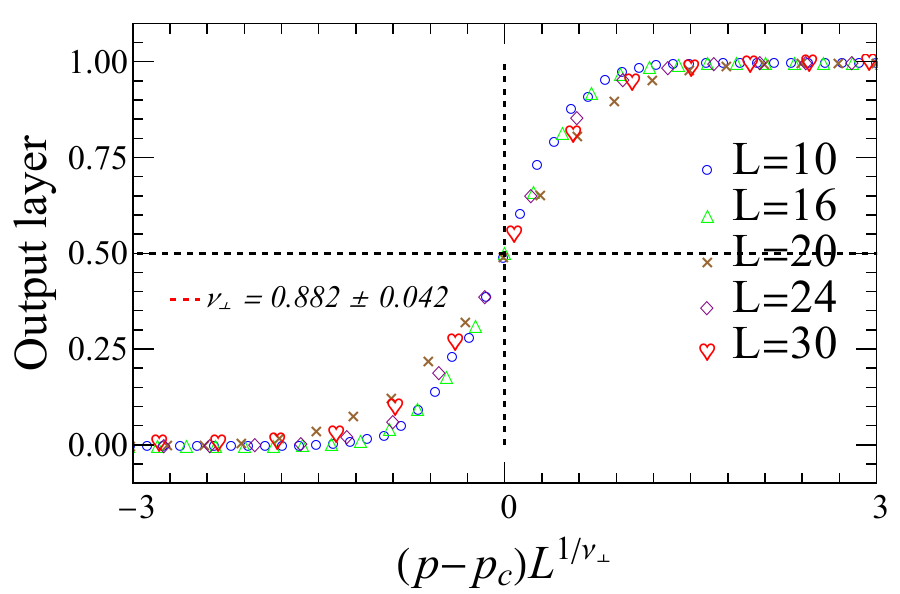}
                \end{minipage}
            }
            \caption{Data collapse analysis for estimating the critical exponent $\nu$ in the 3D site percolation model. Subfigures (a)--(d) correspond to anchor probabilities $p_a=0$, $0.15$, $0.47$, and $1$, respectively.}
            \label{3dsitecollapse}
        \end{minipage}
\end{figure}

\begin{figure}[htbp]
    \centering
        \begin{minipage}{1\linewidth}
            \centering
            \subfigure[$p_a=0$]{
                \begin{minipage}[b]{0.3\linewidth}
                    \centering
                    \includegraphics[width=0.95\textwidth]{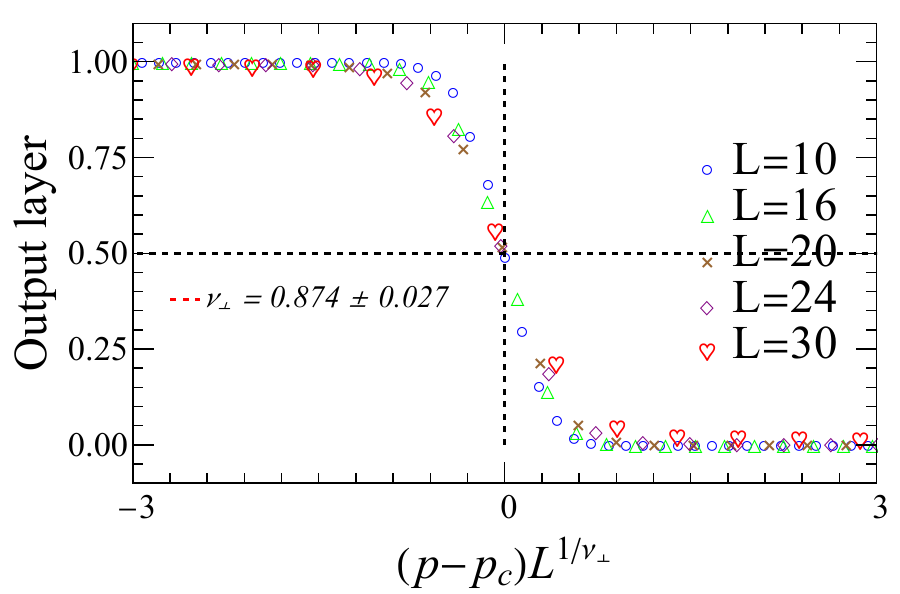}
                \end{minipage}
            }
            \subfigure[$p_a=0.12$]{
                \begin{minipage}[b]{0.3\linewidth}
                    \centering
                    \includegraphics[width=0.95\textwidth]{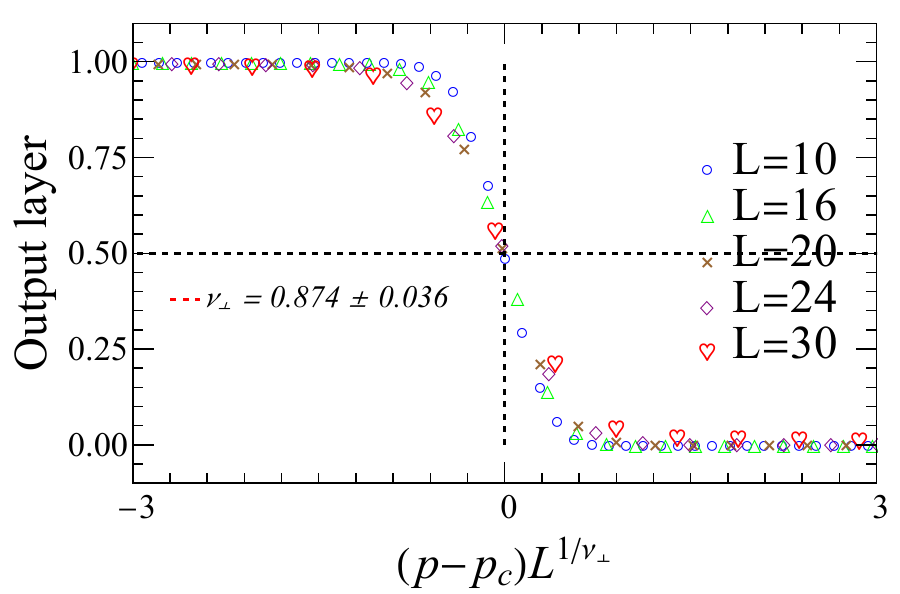}
                \end{minipage}
            }
            \\ 
            \subfigure[$p_a=0.38$]{
                \begin{minipage}[b]{0.3\linewidth}
                    \centering
                    \includegraphics[width=0.95\textwidth]{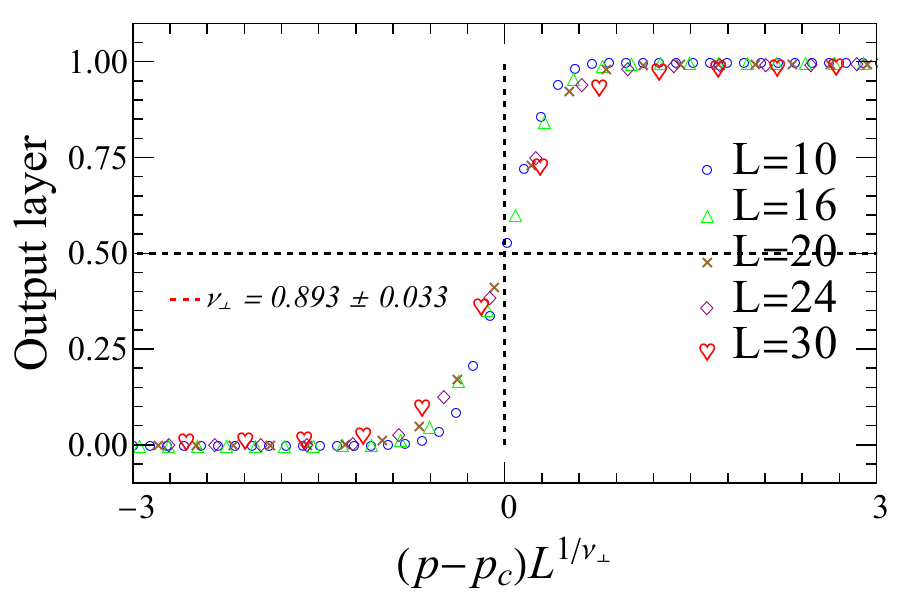}
                \end{minipage}
            }
            \subfigure[$p_a=1$]{
                \begin{minipage}[b]{0.3\linewidth}
                    \centering
                    \includegraphics[width=0.95\textwidth]{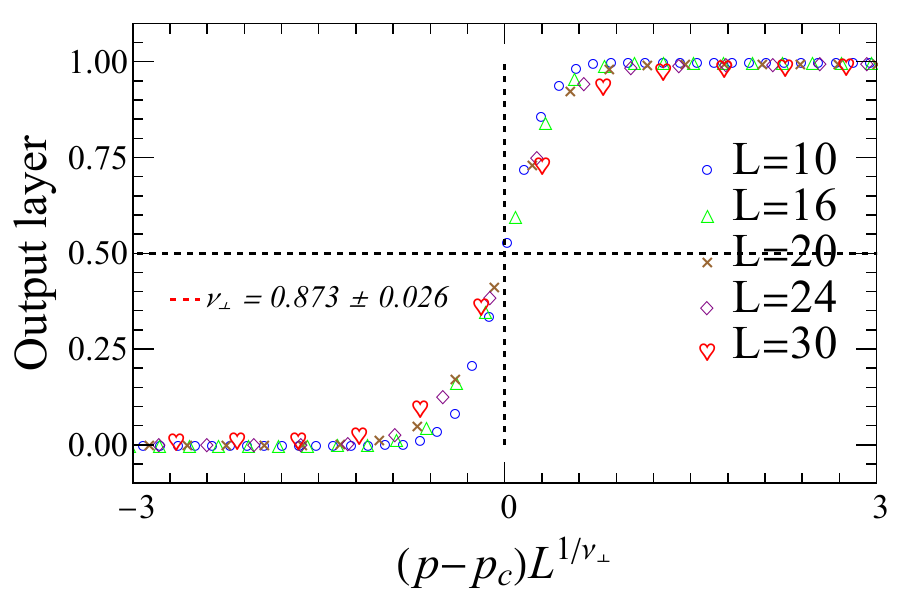}
                \end{minipage}
            }
            \caption{Data collapse analysis for estimating the critical exponent $\nu$ in the 3D bond percolation model. Subfigures (a)--(d) correspond to anchor probabilities $p_a = 0$, 0.12, 0.38, and 1, respectively.}
            \label{3dbondcollapse}
        \end{minipage}
\end{figure}

Furthermore, data collapse analysis was performed on the SNN outputs corresponding to different anchors to assess whether the method can reliably extract the critical exponent $\nu$. The corresponding results are shown in Figures~\ref{3dsitecollapse} and \ref{3dbondcollapse}. 

To quantify the critical exponent $\nu$ and the quality of the data collapse, we defined a cost function $S(\nu)$ based on the mean squared error (MSE) of the residuals relative to a master curve. Specifically, for a given trial $\nu$, we constructed a scaled dataset $\{(x_j, y_j)\}$ where $x_j = (p-p_c)L^{1/\nu}$ and $y_j$ is the SNN output. The cost function is defined as:
\begin{equation}
    S(\nu) = \frac{1}{M} \sum_{j=1}^{M} \left[ y_j - \mathcal{F}_{poly}(x_j) \right]^2,
    \label{eq:cost_function}
\end{equation}
where $M$ is the total number of data points within the restricted window $x_j \in [-3, 3]$, and $\mathcal{F}_{poly}$ denotes a 7th-order polynomial fitted to the scaled data. The optimal exponent $\nu$ was identified by minimizing $S(\nu)$.

To estimate the statistical uncertainty, we employed the Jackknife resampling method over the $N$ available system sizes. The standard error $\sigma_{\nu}$ is calculated as:
\begin{equation}
    \sigma_{\nu} = \sqrt{\frac{N-1}{N} \sum_{i=1}^{N} (\nu_{(i)} - \bar{\nu})^2},
    \label{eq:jackknife}
\end{equation}
where $\nu_{(i)}$ is the estimate obtained by excluding the $i$-th system size, and $\bar{\nu}$ is the average of all jackknife estimates. The quantitative estimates for the critical exponent $\nu$ across all considered anchors are summarized in Table~\ref{tab:nu_values}.

\begin{table}[htbp]
    \centering
        \begin{minipage}{1\linewidth}
            \centering
            \rmfamily 
            \caption{\hl{Critical exponent $\nu$ for 3D percolation by anchor.}}
            \label{tab:nu_values}
            \renewcommand{\arraystretch}{1.2}
            \begin{tabular}{lccc}
                \hline
                Model & Anchor ($p_a$) & Best Fit $\nu$ & Literature~\cite{brzeski2022percolation} \\
                \hline
                \multirow{4}{*}{Site Percolation} 
                 & $0$    & $0.899 \pm 0.031$ & \multirow{4}{*}{$0.8762(7)$} \\
                 & $0.15$ & $0.894 \pm 0.036$ & \\
                 & $0.47$ & $0.882 \pm 0.048$ & \\
                 & $1.0$  & $0.882 \pm 0.042$ & \\
                \hline
                \multirow{4}{*}{Bond Percolation} 
                 & $0$    & $0.874 \pm 0.027$ & \multirow{4}{*}{$0.8762(7)$} \\
                 & $0.12$ & $0.874 \pm 0.036$ & \\
                 & $0.38$ & $0.893 \pm 0.033$ & \\
                 & $1.0$  & $0.873 \pm 0.026$ & \\
                \hline
            \end{tabular}
        \end{minipage}
\end{table}

For the site percolation model, the estimated values range from $0.882$ to $0.899$. For the bond percolation model, the estimates are similarly stable, ranging between $0.873$ and $0.893$. In all cases, the results are consistent with the standard theoretical value $\nu = 0.8762(7)$~\cite{brzeski2022percolation} within the statistical uncertainty. This confirms that the SNN-based feature extraction is robust against anchor selection and reliably captures the universal critical scaling behavior.

\subsection{Extension of Data Labeling Range}
\label{sec44}

\begin{figure}[htbp]
	\centering
	\subfigure[]{
		\begin{minipage}[b]{0.45\textwidth}
			\includegraphics[width=0.9\textwidth]{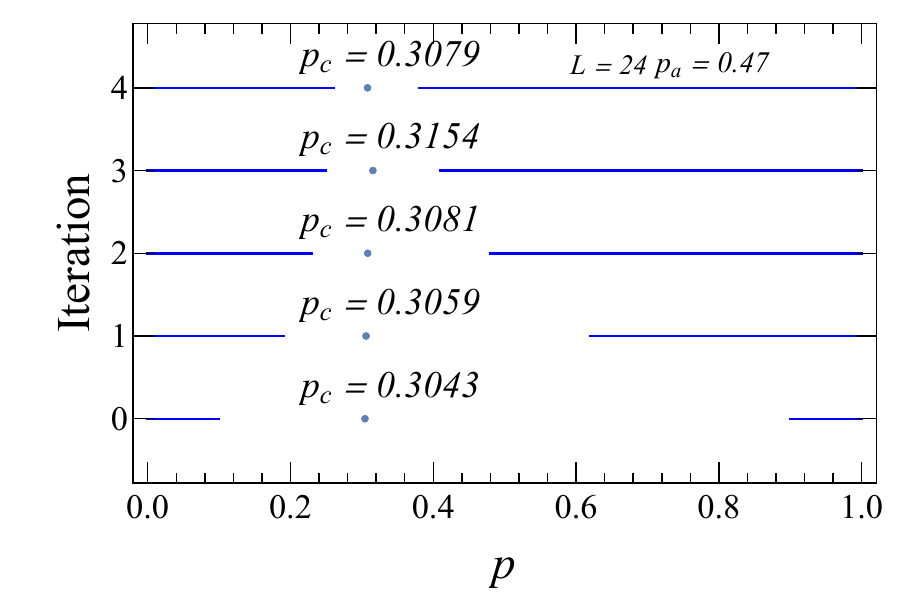} 
		\end{minipage}
	}
	\subfigure[]{
		\begin{minipage}[b]{0.45\textwidth}
			\includegraphics[width=0.9\textwidth]{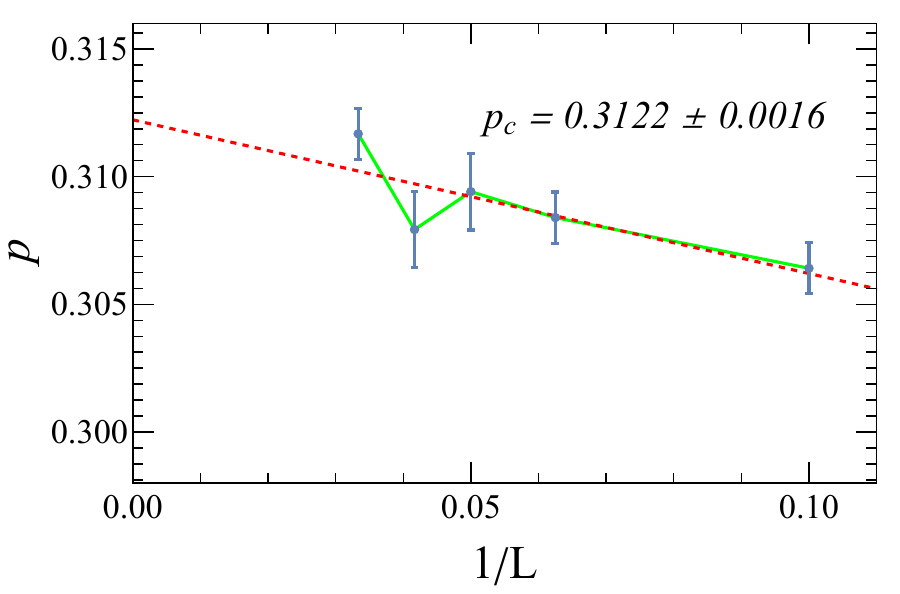}
		\end{minipage}
	}
    \caption{(a) Estimated critical thresholds at anchor \( p_a = 0.47 \) for the three-dimensional site percolation model with system size \( L = 24 \), obtained via iterative adjustment of the probability labeling range. 
    (b) FSS extrapolation at anchor \( p_a = 0.47 \) as a function of \( 1/L \), based on the refined labeling interval.}
    \label{3dsiterange}
\end{figure}

As previously stated, the SNN follows a semi-supervised learning paradigm that requires manual labeling of a subset of the data during training. Inspired by previous work\cite{shenTransferLearningPhase2022} and\cite{Chen2022StudyOP}, we extend the labeling range to study its impact on prediction accuracy. Initially, the interval \( [0, 0.1] \cup [0.9, 1] \) is used as the labeling range, and corresponding model outputs are obtained. Subsequently, the probability values \( l^{(i)} \) and \( r^{(i)} \) are identified as the points at which the similarity curve first falls below 99\% and 1\%, respectively. The initial probability boundaries \( p^{(0)} = 0.1 \) and \( q^{(0)} = 0.9 \) are iteratively updated, and the new labeling interval \( [0, p^{(i)}] \cup [q^{(i)}, 1] \) is computed using Equation~\ref{p1p2}.
\begin{equation} p^{(i)} = \frac{p^{(i-1)} + l^{(i-1)}}{2}, \quad q^{(i)} = \frac{q^{(i-1)} + r^{(i-1)}}{2} \quad(i=1, 2, 3, \dots).
\label{p1p2} 
\end{equation}
The iterative labeling process terminates once the condition \( p^{(i)} = l^{(i)} \) or \( q^{(i)} = r^{(i)} \) is satisfied. Figure~\ref{3dsiterange}(a) illustrates this iterative process for the 3D site percolation model with system size $L=24$ and anchor $p_a = 0.47$. These intervals are selected because configurations within them predominantly belong to a single phase—either fully percolating or entirely non-percolating—and thus exhibit stable, non-critical behavior.

We apply the aforementioned iterative method to determine the optimal labeling intervals for four selected anchors in the three-dimensional site percolation model across various system sizes, as summarized in Table~\ref{3dsiteiteration_range}. For all system sizes, the initial labeling interval is set to \( [0, 0.1] \cup [0.9, 1] \), corresponding to iteration \( i = 0 \). All system sizes require four iterations to converge, except for \( L = 16 \), which converges after three iterations. Upon convergence, the critical thresholds corresponding to different anchors are estimated in the thermodynamic limit. The extrapolation for anchor \( p_a = 0.47 \) is shown in Figure~\ref{3dsiterange}(b), and the results for other anchors are listed in Table~\ref{3dsitepc_v}. Furthermore, a data collapse analysis is performed on the post-iteration data, and the corresponding estimates of the critical exponent \( \nu \) are also reported in Table~\ref{3dsitepc_v}. In this table, the first and third rows present the results obtained before applying the iterative refinement, while the second and fourth rows correspond to the post-iteration outcomes.

\begin{table}
\caption{\hl{Iterative labeling ranges for 3D site percolation.}}
\label{3dsiteiteration_range}
\rmfamily
\begin{tabular*}{\tblwidth}{@{}LLLLL@{}}
\toprule
~ & $i=1$ & $i=2$ & $i=3$ & $i=4$ \\ 
\midrule
$L=10$ & [0,0.15]$\cup$[0.64,1] & [0,0.18]$\cup$[0.51,1] & [0,0.19]$\cup$[0.45,1] & [0,0.20]$\cup$[0.39,1] \\
$L=16$ & [0,0.18]$\cup$[0.62,1] & [0,0.22]$\cup$[0.49,1] & [0,0.24]$\cup$[0.42,1] & ~ \\
$L=20$ & [0,0.18]$\cup$[0.63,1] & [0,0.22]$\cup$[0.49,1] & [0,0.24]$\cup$[0.42,1] & [0,0.25]$\cup$[0.38,1] \\
$L=24$ & [0,0.19]$\cup$[0.62,1] & [0,0.23]$\cup$[0.48,1] & [0,0.25]$\cup$[0.41,1] & [0,0.26]$\cup$[0.38,1] \\
$L=30$ & [0,0.18]$\cup$[0.62,1] & [0,0.22]$\cup$[0.48,1] & [0,0.24]$\cup$[0.41,1] & [0,0.25]$\cup$[0.38,1] \\
\bottomrule
\end{tabular*}
\end{table}

\begin{table}
\caption{\hl{Critical thresholds and exponents for 3D site percolation, before/after iteration.}}
\label{3dsitepc_v}
\rmfamily
\begin{tabular*}{\tblwidth}{@{}LLLLLLL@{}}
\toprule
~ & $p_a=0$ & $p_a=0.15$ & $p_a=0.47$ & $p_a=1$ & MC & Literature\cite{xu2014simultaneous,brzeski2022percolation} \\ 
\midrule
$p_c$         & 0.3103(19) & 0.3090(20) & 0.3149(20) & 0.3147(20) & 0.3146(14) & 0.31160768(15) \\
$p^{it}_c$    & 0.3182(25) & 0.3157(29) & 0.3122(16) & 0.3127(19) & ~ & ~ \\
$\nu$         & 0.88 & 0.88 & 0.88 & 0.88 & 0.88 & 0.8762(7) \\
$\nu^{it}$    & 0.88 & 0.88 & 0.88 & 0.88 & ~ & ~ \\
\bottomrule
\end{tabular*}
\end{table}

Comparison of the FSS and data collapse analyses before and after iteration reveals that extending the labeling interval does not significantly improve the accuracy in predicting either the critical threshold or the critical exponent. For instance, at anchor \( p_a = 0.47 \), the estimated critical threshold improved from \( p_c = 0.3149(20) \) to \( p_c = 0.3122(16) \) after refining the labeling interval. These findings suggest that the critical threshold and critical exponent of the three-dimensional percolation model can be accurately predicted, even with relatively narrow labeling intervals.

\subsection{Generalization and Scalability Analysis}
\label{sec45}

To further assess the robustness of the SNN framework, we conducted additional experiments focusing on two critical aspects: the scalability of the method to larger system sizes and its generalization capability across different lattice geometries.

\textbf{1) Scalability to Larger Systems}

In the previous sections, the maximum system size analyzed was $L=30$. To verify whether the proposed method remains effective for larger systems, we extended the experiments to a SC lattice with $L=64$.
\begin{figure}[htbp]
    \centering
        \begin{minipage}{1\linewidth}
            \centering
            \includegraphics[width=0.4\textwidth]{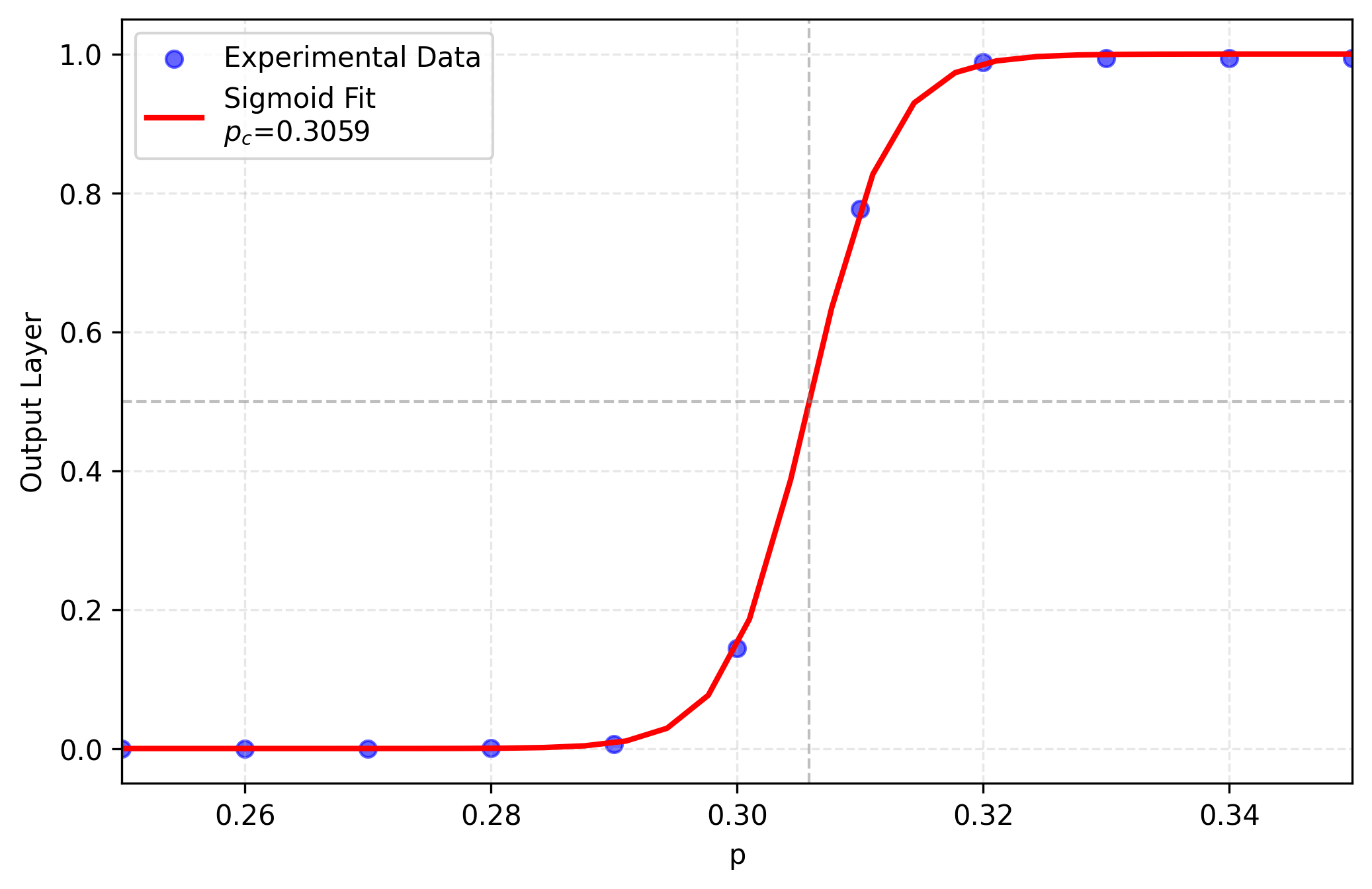}
            
            \caption{Scalability verification on larger systems. The SNN model was retrained on simple cubic lattices with a system size of $L=64$. The blue circles represent the average model output, and the red solid line indicates the sigmoid fit. The estimated critical threshold is $p_c \approx 0.3059$, deviating by approximately $2\%$ from the high-precision Monte Carlo value $p_c = 0.311 607 68(15)$. This experiment serves to verify the scalability of the training procedure to larger system sizes, not to benchmark precision; the achievable precision remains well below that of standard Monte Carlo finite-size scaling analyses.}
            \label{Scalability}
        \end{minipage}
\end{figure}
For this analysis, we generated a dataset containing 101 probability points ($p \in [0, 1]$), with 500 independent configurations per probability. The SNN model was trained on this larger dataset following the same protocol. As shown in Figure~\ref{Scalability}, the model successfully identified the phase transition behavior. 
The estimated critical threshold is $p_c \approx 0.3059$, deviating by approximately $2\%$ from the high-precision Monte Carlo value for SC site percolation, $p_c = 0.311 607 68(15)$.

It is worth noting that due to the use of fully connected layers in our architecture, the input dimension is fixed by the system size ($L^3$). Consequently, a model trained on $L=20$ cannot be directly applied to $L=64$, unlike architectures utilizing global pooling layers. However, the successful prediction at $L=64$ confirms that the methodology itself is scalable and can be readily adapted to larger simulation scales with appropriate retraining.

\textbf{2) Cross-Geometry Generalization}

A key question in machine learning-based physics is whether the model learns intrinsic physical features of the phase transition. To investigate this, we performed a cross-geometry generalization test. Specifically, we took the SNN model trained solely on the SC lattice (coordination number $z=6$) and applied it directly to test configurations generated on a FCC lattice ($z=12$), without any retraining or fine-tuning.

\begin{figure}[htbp]
	\centering
	\subfigure[]{
		\begin{minipage}[b]{0.45\textwidth}
			\includegraphics[width=0.9\textwidth]{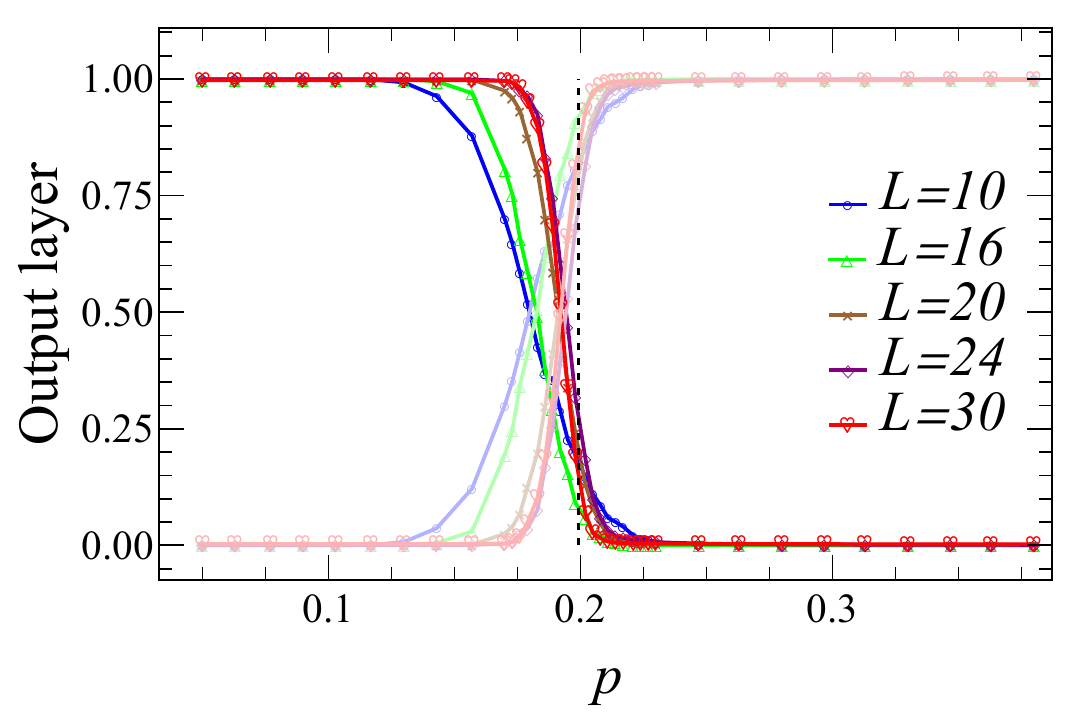} 
		\end{minipage}
	}
	\subfigure[]{
		\begin{minipage}[b]{0.45\textwidth}
			\includegraphics[width=0.9\textwidth]{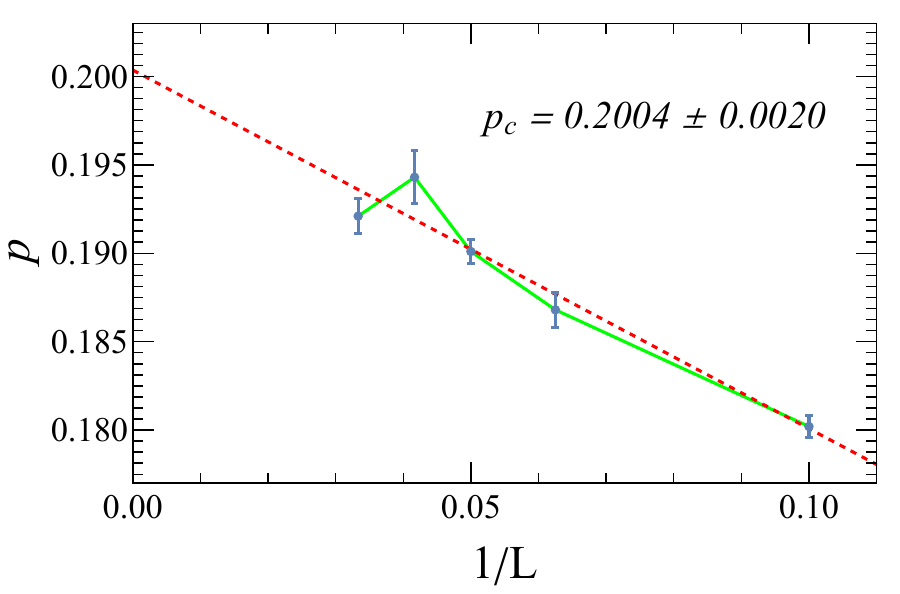}
		\end{minipage}
	}
    \caption{Cross-geometry generalization test on FCC lattices. The SNN model was trained solely on SC lattices and applied directly to FCC lattices without retraining. 
    (a) Average output layer response as a function of occupation probability $p$ for different system sizes $L \in \{10, 16, 20, 24, 30\}$. The intersection of curves indicates the effective identification of the phase transition. 
    (b) FSS analysis. The determined pseudo-critical points $p_c(L)$ are plotted against $1/L$. Linear extrapolation to the thermodynamic limit ($1/L \to 0$, red dashed line) yields a critical threshold of $p_c = 0.2004 \pm 0.0020$, consistent with the theoretical value for FCC percolation, ($p_c^{\text{FCC}} \approx 0.199236(4)$), within one standard error.}
    \label{Generalization}
\end{figure}

We tested FCC systems with sizes $L \in \{10, 16, 20, 24, 30\}$. As illustrated in Figure~\ref{Generalization}(a), despite the topological difference between SC and FCC lattices, the SNN output correctly reflects the order parameter behavior.

We further performed a FSS analysis on the predicted critical points $p_c(L)$ derived from the FCC data. As shown in Figure~\ref{Generalization}(b), the extrapolation to the thermodynamic limit yields a critical threshold of $p_c \approx 0.2004$. This result is consistent with the theoretical threshold for FCC site percolation ($p_c = 0.199236(4)$)\cite{hu2021percolation}within statistical uncertainty. The result confirms the cross-geometry generalization capability of the proposed SNN framework.

\subsection{Analysis of Learned Representations}
\label{sec46}

While the SNN demonstrates high accuracy in locating critical thresholds, a fundamental question remains: does the model rely on physically meaningful features or merely exploit spurious statistical correlations? To investigate this, we analyzed the internal representations learned by the network. It is important to note that the input to our network is the DFS-extracted largest cluster; therefore, we expect the learned features to reflect structural properties of the largest cluster.

\begin{figure}[htbp]
    \centering
        \begin{minipage}{1\linewidth}
            \centering 
            \includegraphics[width=0.95\textwidth]{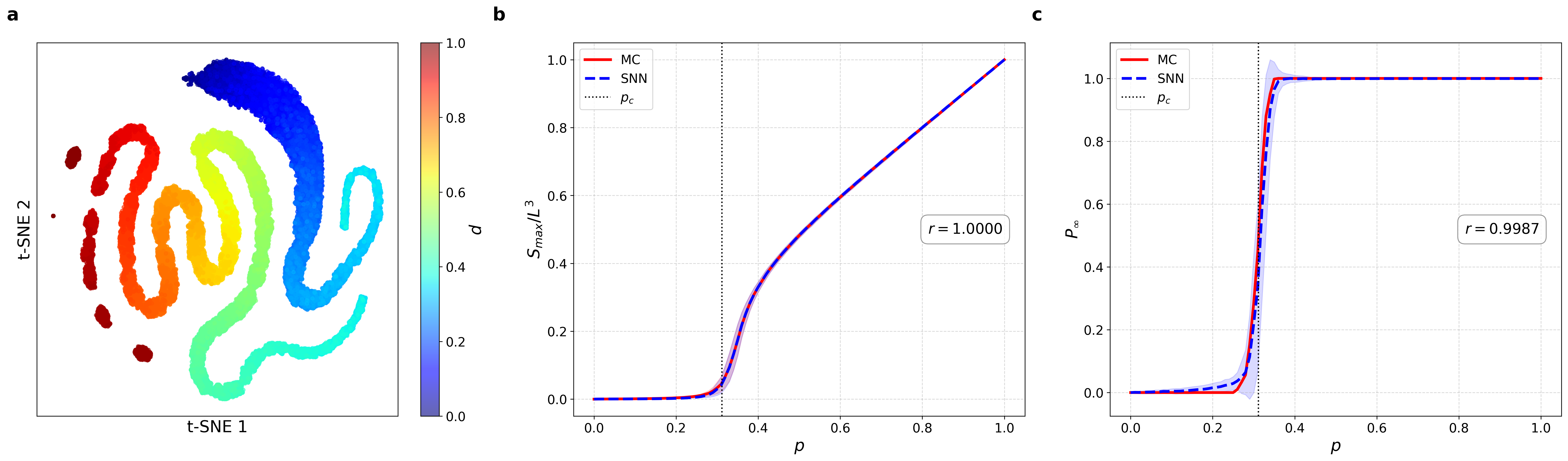}
            \caption{Visualization and analysis of the SNN learned representations for system size $L=20$. (a) t-SNE projection of the high-dimensional embeddings. Points are colored by their occupation probability $p$, revealing a structured manifold ordered by $p$ despite the absence of regression supervision. (b) Correlation between the SNN's latent features and the normalized largest cluster size. The first principal component (PC1) of the embedding (blue dashed line) exhibits a strong correlation ($r > 0.99$) with the normalized largest cluster size $S_{max}/L^3$ obtained via Monte Carlo simulations (red solid line). (c) Comparison between the SNN similarity output and the order parameter. The SNN score relative to a percolating anchor (blue dashed line) closely tracks the percolation probability $P_{\infty}$ (red solid line), demonstrating that the model effectively captures the macroscopic transition behavior.}
            \label{fig:interpretability}
        \end{minipage}
\end{figure}

We extracted the 32-dimensional embedding vectors from the feature extractor and projected them into a two-dimensional space using t-Distributed Stochastic Neighbor Embedding (t-SNE). As visualized in Figure.~\ref{fig:interpretability}(a), the configurations map onto a continuous manifold structure. Although the network was trained solely on pairwise similarity labels (0 or 1) without explicit knowledge of the occupation probability $p$, the projected points exhibit an ordered arrangement with respect to $p$. The separation between the non-percolating phase (blue/green region) and the percolating phase (orange/red region) suggests that the SNN has learned a metric space where structural changes associated with the phase transition are mapped to spatial proximity.

To quantitatively assess the physical relevance of these embeddings, we examined their relationship with standard order parameters. Figure~\ref{fig:interpretability}(b) plots the first principal component (PC1) of the SNN embeddings against the normalized size of the largest cluster ($S_{max}/L^3$), which serves as the order parameter for percolation. The two quantities exhibit a strong linear correlation ($r > 0.99$). This indicates that the primary variance in the SNN's latent space effectively encodes the mass of the largest cluster.

Furthermore, Figures~\ref{fig:interpretability}(c) compares the SNN's output similarity score (measured against a reference anchor at $p=1$) with the percolation probability $P_{\infty}$ (the probability that a site belongs to the infinite cluster). The SNN output reproduces the functional form of $P_{\infty}$ with high fidelity ($r > 0.99$).

Why does the network converge to this particular representation? Two ingredients combine. First, the training task--discriminating largest clusters drawn from $p \in [0, 0.1]$ versus $p \in [0.9, 1]$---is solvable, with a large margin, by a single scalar statistic: the mass of the input cluster, which differs by orders of magnitude between the two regimes. Second, fully connected layers access such globally summed statistics most readily, and gradient-based training is known to favor the simplest features sufficient for the task. The raw-input ablation in Section~\ref{sec43} provides direct evidence of this inductive bias: fed unprocessed configurations, the identical architecture learns the global occupation density $\rho = p$, producing an incorrect crossover at $p \approx 0.22$.

This observation allows the role of the DFS preprocessing to be stated more sharply: it aligns the network's inductive bias with the physics. For the largest-cluster tensor, the ``density'' that the network naturally extracts is no longer the trivial $\rho = p$ but $S_{max}/L^3$---precisely the finite-size order parameter of percolation. The preprocessing thus converts the network's bias toward density-like statistics from a failure mode (raw input, Fig.~6) into the mechanism of success (cluster input, Fig.~14b). The strong correlation between the first principal component and $S_{max}/L^3$ ($r > 0.99$) is the quantitative signature of this alignment.

Within this picture, threshold detection itself acquires a clear physical interpretation: the similarity score against a fixed anchor is a monotone readout of the learned mass-like feature, and therefore changes most rapidly where $S_{max}(p)$ does---namely in the critical region---yielding the sigmoidal response whose midpoint defines $p_c(L)$. The same picture accounts for the cross-geometry transfer of Section~\ref{sec45}: what the network encodes is an intrinsic property of the cluster---its mass---rather than of the embedding lattice, and the learned metric therefore remains meaningful when the coordination number changes from $z = 6$ (SC) to $z = 12$ (FCC).

A natural question follows: if the network essentially learns an $S_{max}$-like quantity, why not measure $S_{max}$ directly? The point is that the network arrives at this quantity without being told which observable to compute---it is supplied only with pairwise similarity labels. For regular lattice percolation, where the order parameter is textbook knowledge, this autonomy yields no practical advantage, and the present results on SC and FCC lattices serve precisely to verify that the autonomously discovered feature does coincide with the established order parameter. This verification, achievable only on systems with known answers, is what motivates the framework's potential extension to systems where a quantitative order parameter is not explicitly defined but a relevant structural motif can still be identified---a direction we leave to future work.

Finally, we note the limits of this analysis. Principal component analysis captures only the dominant direction of variance; the remaining embedding dimensions may encode additional geometric information (e.g., cluster shape or surface structure), which plausibly contributes to the transferability but remains to be characterized.

\section{Conclusion}

In this study, we systematically investigated the application of SNN to site and bond percolation models defined on a three-dimensional cubic lattice. 
By treating the extracted largest cluster as a structured input, our framework learns to map topological similarities into a latent metric space. The results indicate that the SNN approach, previously applied to two-dimensional systems, can be effectively extended to three-dimensional models, identifying phase transitions and estimating critical exponents with reasonable accuracy despite the increased topological complexity.

A notable observation from our experiments is the cross-geometry transferability: a model trained solely on SC lattices ($z=6$) identifies the phase transition in FCC lattices ($z=12$) without retraining. This capability is quantitatively grounded in the representation analysis of Section~\ref{sec46}: the learned embedding predominantly encodes the mass of the largest cluster---an intrinsic property of the cluster rather than of the embedding lattice---which explains why the similarity metric survives the change in coordination number. More broadly, the mechanistic picture established there shows that the network, given only weak pairwise supervision and a generic structural prior, autonomously converges to a statistic that coincides quantitatively with the percolation order parameter.

We emphasize that the proposed framework is not positioned as a competitor to Monte Carlo methods. For regular lattice percolation, MC simulations combined with finite-size scaling remain the gold standard, achieving precision ($\sim 10^{-7}$) that the present approach does not approach ($\sim 10^{-2}$). The two methodologies instead occupy complementary roles. MC+FSS presupposes knowledge of which scalar observable to measure and benefits from dense sampling near criticality---requirements that are entirely natural for well-characterized lattice models. The SNN framework relaxes precisely these requirements: it needs no quantitative order parameter (only a generic connectivity prior), no labels near the critical region, and---uniquely---transfers across lattice geometries without retraining, a capability with no analogue in the MC paradigm. These properties offer no advantage on regular lattices; they become relevant in settings where the order parameter is not explicitly defined or data acquisition is constrained, such as experimental imaging data or complex disordered media. A remaining limitation is that the choice of connectivity as the relevant structural motif is itself informed by percolation theory; for systems where the relevant motif is unknown, the preprocessing step becomes part of the discovery problem, and exploring more generic inputs (e.g., full cluster-size distributions or topological summaries) is left to future work.

From an application perspective, the SNN-based approach exhibits practical potential across domains involving critical phenomena, such as estimating transition thresholds in material systems~\cite{xie2018crystal, peixoto2019bayesian}, identifying structural stability in complex networks, and analyzing geophysical phenomena such as rock fracture and fluid seepage~\cite{wang2021deep}.

In summary, we propose a label-efficient learning framework for identifying phase transitions in three-dimensional percolation models. Its distinguishing value lies in the combination of three properties absent from the MC paradigm ---requiring no quantitative order parameter, learning exclusively from labels far from criticality, and transferring across lattice geometries---together with an evidence-based account of the representation the network learns.

\section*{Acknowledgments}
This work was supported in part by the National Key Research and Development Program of China (Grant No. 2024YFA1611003), the Fundamental Research Funds for the Central Universities (XJ2026002701), the Natural Science Foundation of Fujian Province (Grant No.2026J0011623), the Research Fund of Baoshan University (BYKY202305); Yunnan Fundamental Research Projects (Grant No. 202401AU070035); the Fundamental Research Funds for the Central Universities, China (Grant No. CCNU19QN029); the National Natural Science Foundation of China (Grant No. 61873104); the 111 Project 2.0 (Grant No. BP0820038); and the self-determined research funds of CCNU from the colleges’ basic research and operation of MOE.

\clearpage

\printcredits

\bibliographystyle{elsarticle-num-names}

\bibliography{sample}

\end{document}